\documentclass[journal]{IEEEtran}
\usepackage{xcolor,soul,framed} 

\usepackage{physics}
\usepackage{url}
\usepackage[utf8]{inputenc}
\usepackage{color}
\usepackage{graphicx}
\usepackage{graphics} 
\usepackage{epsfig} 
\usepackage{mathptmx} 
\usepackage{times} 
\usepackage{amsmath} 
\usepackage{amssymb}  
\usepackage{float} 
\usepackage{subfigure}
\usepackage{multirow}
\usepackage{pifont }
\usepackage{url}
\usepackage{stfloats}
\usepackage{float}
\usepackage{enumerate}
\usepackage{amsfonts}
\usepackage{cite}
\usepackage{float} 
\usepackage{subfigure}
\usepackage{multirow}
\usepackage{pifont }
\usepackage{enumerate}
\usepackage{flexisym}
\usepackage{amsfonts, diagbox}
\usepackage{cite}
\usepackage{mathrsfs}
\usepackage{algorithm}
\usepackage{algorithmic}
\begin{document}
\bstctlcite{IEEEexample:BSTcontrol}
    \title{Robotic Inspection of Underground Utilities for Construction Survey Using a Ground Penetrating Radar}

\author{Jinglun Feng$^{1}$, Liang Yang$^{1}$, Ejup Hoxha$^{1}$, Jiang Biao$^{2}$, Jizhong Xiao$^{1*}$~\IEEEmembership{Senior Member,~IEEE}
\thanks{$^{1}$The CCNY Robotics Lab, Electrical Engineering Department,
The City College of New York, New York, USA
        email(jfeng1,lyang1,ehoxha,jxiao@ccny.cuny.edu)}
\thanks{$^{2}$Hostos Community College, New York, NY, USA. email(bjiang@ccny.cuny.edu)}
\thanks{$^{*}$Corresponding author.}
}

\maketitle

\begin{abstract}
Ground Penetrating Radar (GPR) is a very useful non-destructive evaluation (NDE) device for locating and mapping underground assets prior to digging and trenching efforts in construction. This paper presents a novel robotic system to automate the GPR data collection process, localize the underground utilities, interpret and reconstruct the underground objects for better visualization allowing regular non-professional users understand the survey results. This system is composed of three modules: 1) an  Omni-directional robotic data collection platform, that carries a RGB-D camera with Inertial Measurement Unit (IMU) and a GPR antenna to perform automatic GPR data collection, and tag each GPR measurement with visual positioning information at every sampling step; 2) a learning-based migration module to interpret the raw GPR B-scan image into a 2D cross-section model of objects; 3) a 3D reconstruction module, i.e., GPRNet, to generate underground utility model represented as fine 3D point cloud. Comparative studies are performed on synthetic data and field GPR raw data with various incompleteness and noise. Experimental results demonstrate that our proposed method achieves a $30.0\%$ higher GPR imaging accuracy in mean Intersection Over Union (IoU) than the conventional back projection (BP) migration approach, and $6.9\%$-$7.2\%$ less loss in Chamfer Distance (CD) than baseline methods regarding to point cloud model reconstruction. The GPR-based robotic inspection provides an effective tool for civil engineers to detect and survey underground utilities before construction.

\end{abstract}


%
\IEEEpeerreviewmaketitle



\section{Introduction}

\IEEEPARstart{G}{round} Penetrating Radar (GPR) is widely used in non-destructive testing (NDT) for civil engineers to locate and map buried objects (e.g., utilities, rebars, underground storage tanks, metallic or plastic conduits), measure pavement thickness and properties, locate and characterize subsurface features (e.g., subgrade voids below concrete slabs or behind retaining walls). The GPR inspection is a wave propagation technique that transmits a pulse of polarized high-frequency electromagnetic (EM) waves into the subsurface media. EM wave attenuates as it travels in media and reflects when it encounters a material change. GPR antenna would thus record the strength and traveled time of each reflected pulse. \cite{demirci2012ground}. The reflections are then amplified, processed, and displayed on a screen as an A-scan signal, analogous to a waveform in an oscilloscope. When the GPR device moves along a straight line perpendicular to utility pipes, the ensemble of the A-scans forms a B-scan, which is shown as the hyperbolic feature, indicating the objects' locations as well. 

There are two pain-points limiting the GPR applications on subsurface flaws reveal and underground objects reconstruction. The first one is how to determine GPR’s position and orientation accurately and in real time, and synchronize with GPR measurements at each GPR sampling step. In current practice of GPR data collection, inspectors would either move a GPR cart along pre-marked grid lines and count on the survey wheel encoder to trigger GPR measurements while obtain the accurate linear positions for detailed mapping and survey, or count on high-precision global positioning system (GPS) to provide accurate position information for detecting large underground objects or surveying a large area along nonlinear trajectories. 

In encoder-triggered manual data collection, it is time-consuming and tedious for the inspector to pre-mark the grid intersection points, take notes and photographs, and push the GPR device to closely follow the gird lines in X-Y directions. 
On the other hand, GPS equipment is expensive while its accuracy is still not sufficient for the 3D GPR imaging projects where every scan must be accurately localized and targets must be resolved in inches. In addition, GPS accuracy is degraded in urban environments where buildings may obstruct and distort GPS signals \cite{wells1987guide}. Furthermore, GPS cannot work in indoor environments.



The second challenge is how to develop an efficient 3D GPR imaging method to visualize the subsurface objects allowing regular non-professional people to understand. Unfortunately, the current commercial GPR post-processing software cannot process GPR data collected from non-linear trajectories.



To address these challenges, we implement a low-cost vision-based positioning method, tag the pose information at each GPR sample, and develop GPR analysis software that provides a holistic solution for automated GPR data collection and 3D GPR imaging and reconstruction. 
\begin{figure*}
    \centering
    \includegraphics[width=0.9\textwidth]{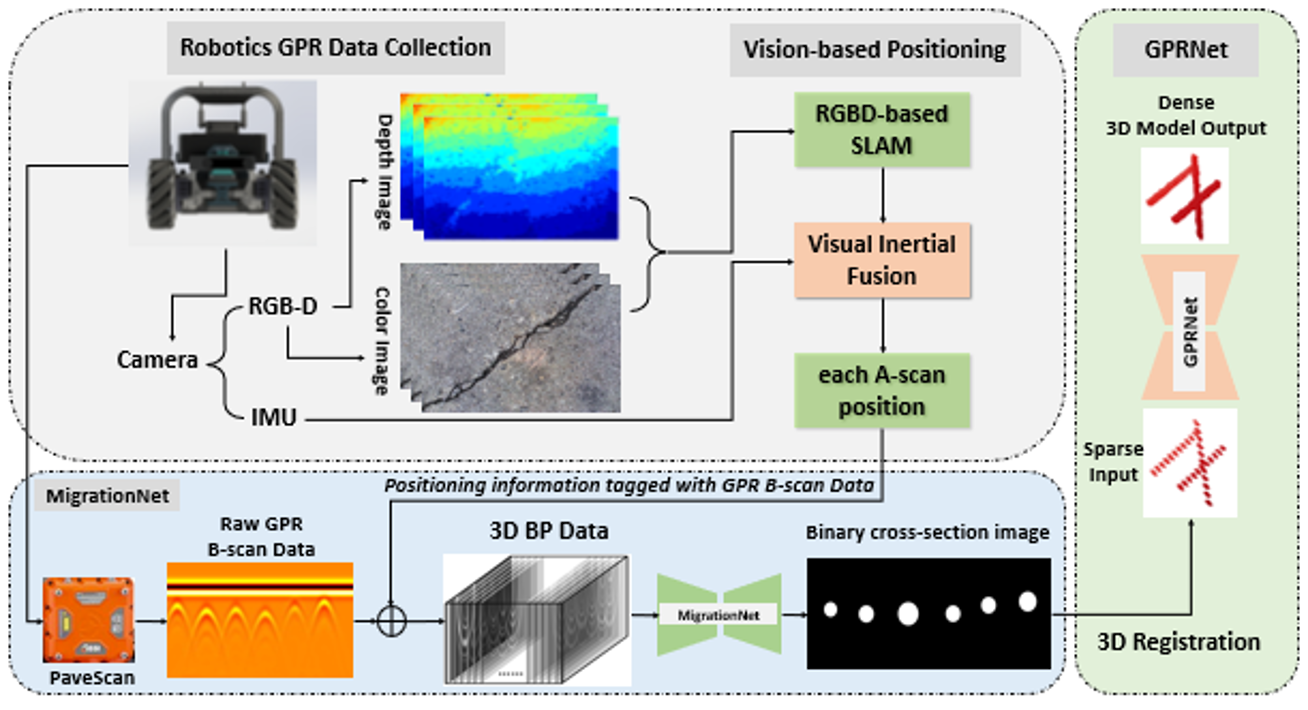}
    \caption{\textbf{System Architecture.} This system contains three modules. First, a vision-based robotic GPR data collection module. It automates the GPR data collection and tags the GPR data with visual positioning information. Second, a MigrationNet to interpret the B-scan image to a 2D cross-section image of the object model. Third, a GPRNet to register 2D cross-section images into the 3D space as a sparse model, and transfer the sparse model to the 3D dense model of subsurface objects.}
    \label{fig:robotics_system}
\end{figure*}
As shown in Figure~\ref{fig:robotics_system}, the main contributions of this work can be summarized as follows:
\begin{itemize}
    \item an Omni-directional robot that carries a GPR antenna on the chassis, moves forward, backward, and sideways in a fast and swift manner and enables the detailed survey to be conducted in arbitrary nonlinear trajectories.
    \item a learning-based module that two separate Deep Neural Networks (DNN). The first network aims to interpret the B-scan data while the second network reveals the object model structure based on the interpreted data and presents the 3D model in point cloud. 
    \item Compared with our previous work \cite{feng2021gpr}, we further augment our GPR dataset with the point cloud models. We released this unique dataset ~\footnote{\url{https://www.dropbox.com/sh/64ermmjrng0dsp2/AABNF2WBvOhV2GO7vmwGEeHUa?dl=0}} to contribute to the learning-based GPR processing communities.
\end{itemize}

This paper is organized as follows. In Section~\ref{section:related_works}, we review related works on conventional GPR migration methods, as well as the recent GPR 3D reconstruction algorithms. Section~\ref{section:data_collection} introduces the robotic GPR data collection system and verifies the vision-based positioning accuracy. Section~\ref{section:migrationnet} and Section~\ref{section:GPRNET} present the learning-based GPR data processing methods. Section~\ref{section:experiment} introduces the proposed GPR dataset and presents extensive experimental results. At last, Section~\ref{section:conclusion} concludes the paper and presents future research directions.

\section{Related Works}
\label{section:related_works}
Ground Penetrating Radar (GPR) migration and model reconstruction are popular topics in NDT and civil engineering and have been extensively investigated in past two decades. 




\textbf{Conventional GPR Migration Methods}:
GPR \emph{migration} is a process that converts the unfocused raw B-scan radargram data into a focused target. Conventional migration methods can be roughly categorized into Kirchhoff migration\cite{schneider1978integral}, the phase-shift migration\cite{gazdag1978wave}, the finite-difference method\cite{claerbout1972downward}, and back-projection (BP) algorithm. 

BP algorithm is the most significant and commonly used 2D imaging reconstruction method in GPR industry \cite{demirci2012study,demirci2012ground}. When GPR emits the radiation pulse, the BP algorithm assumes this wave path shares a semi-sphere pattern with an equal energy level. After GPR receives the radiation pulse back, the BP algorithm stack the radiation energy along the hyperbolic trajectory, then the sum of the responded amplitude could reflect the target region \cite{schofield2020image, gonzalez2014comparative}. 

To further improve the effectiveness of the conventional BP algorithm, several modified BP methods are proposed. X. Xie \emph{et al.} \cite{xie121back} presented the bi-frequency BP (BBP) to enhance the visualization quality of the subsurface objects, especially for grouting. Fast BP (FBP) is proposed by L. Zhou \emph{et al.} \cite{zhou2011fast}, it's an approximation method which could run faster by simplifying the computation of subsurface dielectric \cite{gharamohammadi2019imaging}. In addition, many researches focus on cross-correlation BP (CBP) method \cite{cai2020cross,lin2020forward, jacobsen2010improved, zhang2015back} as well, CBP can cut down the round trip time-of-flight from a stimulating source to a focal point and back to a receiver. Moreover, H. Liu \emph{et al.} \cite{liu2020migration} improved BP algorithm by integrating a correction factor for radiation pattern in the subsurface, to reduce the negative influence of the traditional homogeneous radiation pattern on GPR. Filtered BP (FBP) is another modified BP method, papers \cite{schofield2020image, chetih2015tomographic} investigated this method to get rid of the noise effects back in the GPR images. 

\textbf{3D GPR imaging methods}:
In recent years, research on GPR imaging has made commendable progress in academia \cite{dinh2021full, hou2021improved, qin2021automatic, xiang2021robust}. Prof. Dezhen Song’s group at Taxes A\&M University published a series of papers on automatic subsurface pipeline mapping and 3D reconstruction using a GPR and a camera  \cite{li2019toward,chou2016extrinsic,chou2017mirror,chou2018encoder,li2018robotic,chou2020encoder}. They model the GPR sensing process and prove hyperbola response for general scanning with non-perpendicular angles, which is novel. They fuse V-SLAM and encoder readings with GPR scans to classify hyperbolas into different pipeline groups and apply the J-linkage and maximum likelihood method to estimate the radii and locations of all pipelines. However, the average error for pipe radius estimation is over 35\%, which is not good enough for practical use \cite{li2019toward}. They encounter calibration and synchronization problems and have to use customized artificial landmarks to synchronize camera poses (temporally evenly-spaced) to the GPR data (spatially evenly spaced) \cite{chou2020encoder}. 

Similarly, M. Pereira \emph{et al.} at the University of Vermont published several papers related to 3D reconstruction from both ground and air-coupled multistatic GPR \cite{pereira20183d,pereira20193,pereira20193d,pereira20203}. The main contribution of these works is the consideration of phase compensation for different receiver antennas. They not only stack the B-scan images to model the 3D multistatic GPR imaging, but also take the different gains and dielectric contrast of each receiver antenna into consideration and further fuse it with a Hessian-based enhancement filter to formulate the final 3D reconstruction model. However, the noise reconstructed in the 3D model is still not clearly removed by the proposed method, which makes the 3D model not good enough for visualization. In addition, the author utilized a Google Tango device to provide position information to GPR scan data \cite{pereira2018new}, however, the limitation of this method is that Google Tango is no longer in service and thus this method cannot be implemented in practice. 
\section{Vision-aided Robotic GPR Data Collection}

\label{section:data_collection}
\subsection{Robotic Data Collection Platform}

\begin{figure}
    \centering
    \includegraphics[width=0.45\textwidth]{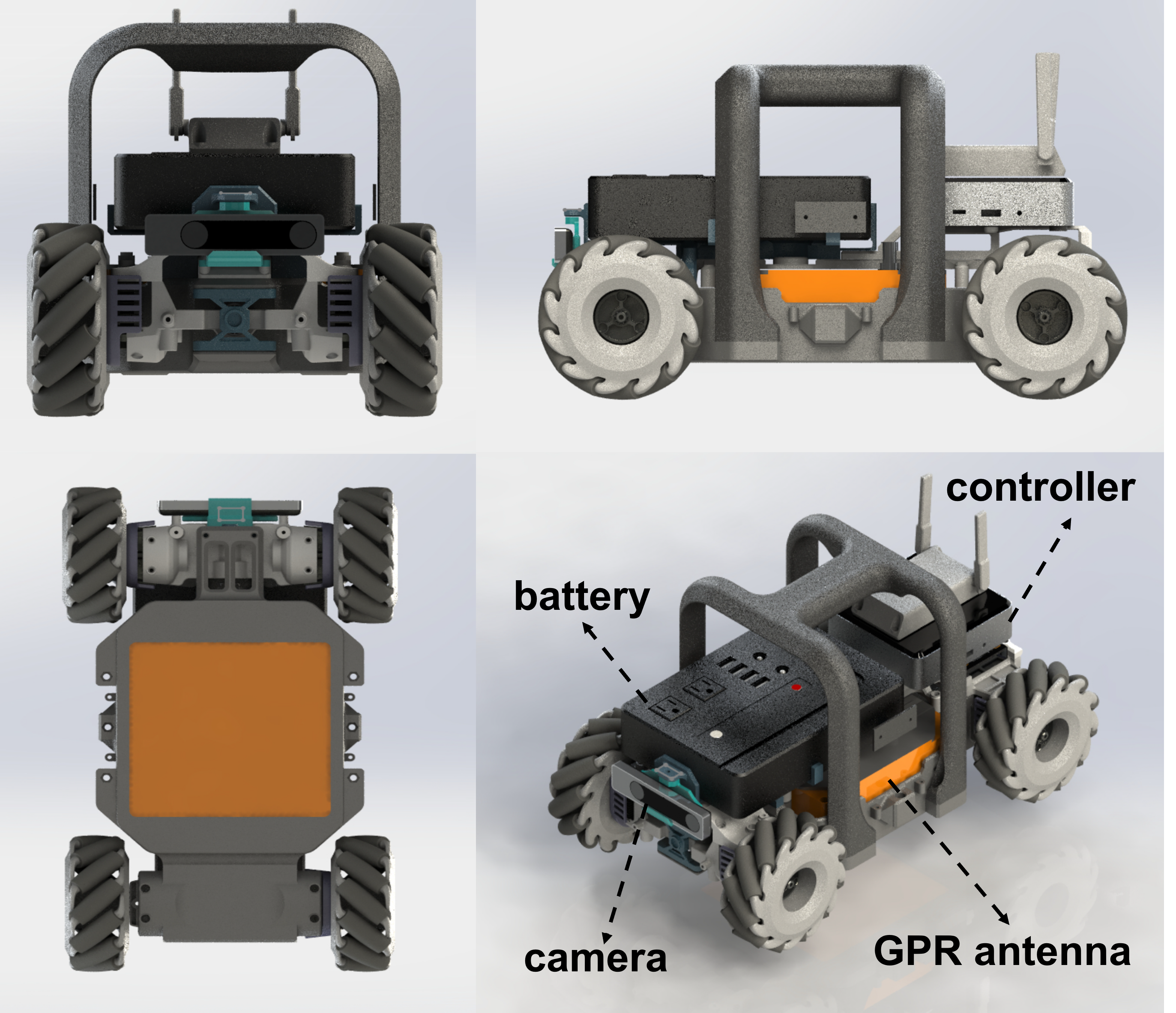}
    \caption{The Omni-directional robot for vision-based GPR data collection, where a GPR antenna is installed at the bottom of the robot chassis.}
    \label{fig:omni_robot}
\end{figure}

As shown in Figure~\ref{fig:omni_robot}, we developed an Omni-directional robot for the inspection of underground utilities. Our robot has four Mecanum wheels that allow forward, backward, and sideways motion to follow grid pattern without spinning. A PaveScan GPR antenna from Geophysical Survey System Inc. (GSSI) is installed at the bottom of the robot chassis to perform GPR data collection. An Intel Realsense RGB-D camera (D435i) is mounted at the robot's front. This camera could support indoor and outdoor working environment, which boosts the robustness for our vision-based positioning system. Note that a 6-axis IMU is embedded in the camera to provide accurate and robust pose estimation. The robot carries a rechargeable battery and a high-level controller (i.e., Intel NUC) to provide power to the system and synchronize the pose data with GPR scan data.   

\begin{figure}[H]
    \centering
    \includegraphics[width=0.3\textwidth]{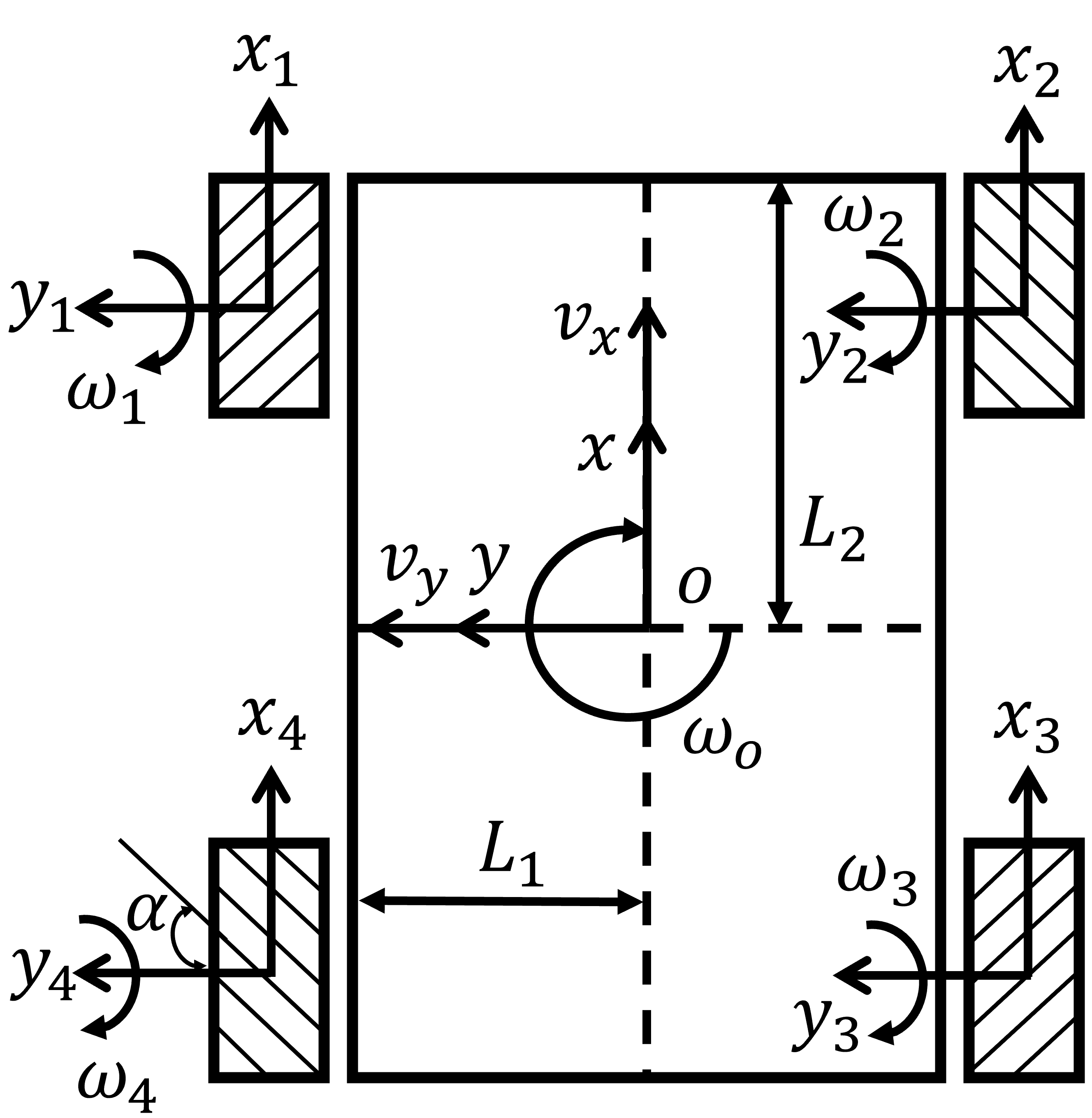}
    \caption{Structural schematic diagram of our GPR cart.}
    \label{fig:mechnum_model}
\end{figure}

Figure~\ref{fig:mechnum_model} depicts the inverse kinematics model of the omni-directional robot. The high maneuverable design allows the robot to move in any directions without spinning and thus provide free motion trajectories for the GPR data collection. The robot motion satisfies the following equation:

\begin{equation}
\begin{aligned}
v_{x}=\frac{R}{4}\left(w_{1}+w_{2}+w_{3}+w_{4}\right) \\
v_{y}=\frac{R}{4}\left(-w_{1}+w_{2}-w_{3}+w_{4}\right) \\
\omega_{o}=\frac{R}{4(L_{2} \tan \alpha+L_{1})}\left(-w_{1}+w_{2}+w_{3}-w_{4}\right)
\end{aligned}
\label{equ:mechnum_model}
\end{equation}

where $R$ is the radius of the Mecanum wheel, $\left\{w_{i}\right\}_{i=1}^{N=4}$ indicates the angular velocity of each Mecanum wheel, while $L_1$ and $L_2$ represents the width and length of the robot chassis respectively. Note that $\alpha$ is the angle of the roller, which equals to $45\textdegree$. $v_{x}$, $v_{y}$ and $\omega_{o}$ are the linear velocity in $x$ and $y$ direction and the angular velocity of the robot chassis respectively. Note that $\left\{x_{i}\right\}_{i=1}^{N=4}$ and $\left\{y_{i}\right\}_{i=1}^{N=4}$ represent the local coordinate frame of each Mecanum wheel, while $x$, $y$ represent the robot coordinate frame of the robot motion, where $o$ denotes the center of the robot chassis and the robot coordinate origin.

Equ~\ref{equ:antenna} demonstrates how robot position and orientation update. In detail, the current pose [$x_{t+1}$, $y_{t+1}$, $\theta_{t+1}$] is updated according to its pose information [$x_{t}$, $y_{t}$, $\theta_{t}$] at previous time, as well as the robot orientation angel $\theta$ and the unit sampling time $\Delta t$. 

\begin{equation}
\begin{aligned}
&\theta_{t+1}=\theta_{t}+\omega_{o} * \Delta t \\
&x_{t+1}=x_{t}+v_{x} \cos \theta_{t+1} * \Delta t-v_{y} \sin \theta_{t+1} * \Delta t \\
&y_{t+1}=y_{t}+v_{x} \sin \theta_{t+1} * \Delta t+v_{y} \cos \theta_{t+1} * \Delta t
\end{aligned}
\label{equ:antenna}
\end{equation}

\begin{figure}[H]
    \centering
    \includegraphics[width=0.45\textwidth]{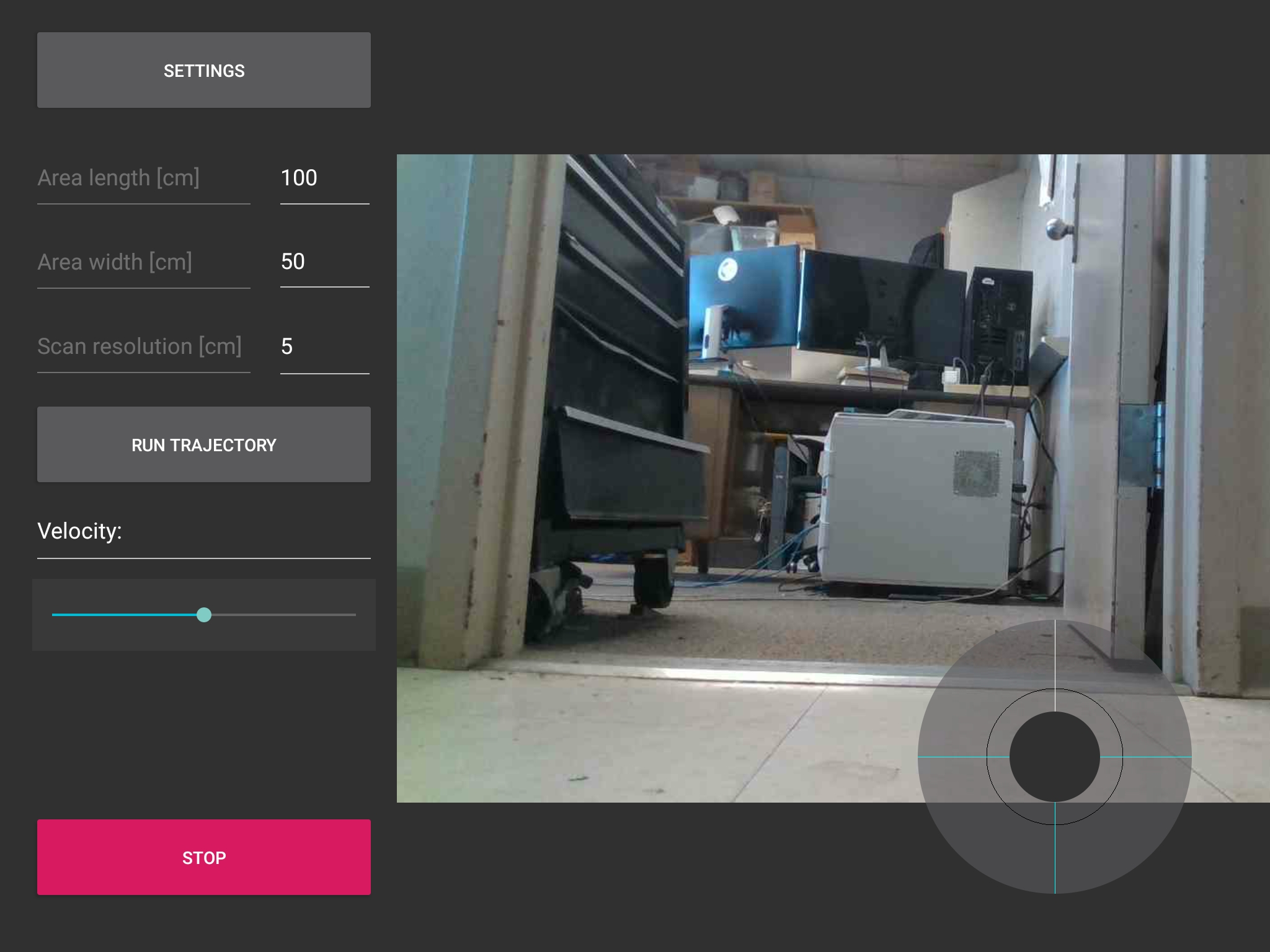}
    \caption{Remote controller App GUI of our robot data collection platform showing First Person View}
    \label{fig:app_view}
\end{figure}

To remote control the robot motion and conduct GPR data collection, we design an Android APP whose graphic user interface (GUI) is illustrated in Figure~\ref{fig:app_view}. We provide two control modes in this remote control APP: 1) automatic and 2) manual. In automatic mode, the user needs to define the width, length and grid resolution of the survey area, then our APP will generate a zig-zag path along the grid to cover the survey area according to the pre-defined parameters, and the robot start the data collection automatically. Note that the user could stop an automatic data collection by pressing the red stop button at bottom left of the GUI. On the other hand, manual mode can be activated anytime to control the robot with the virtual joystick button (the double circles at the bottom right of the GUI). The virtual joystick button controls the robot motion direction by pushing the middle gray circle to the desired angle. If joystick is released and the gray circle rest in the middle then robot stops. Moreover, the user could define the linear velocity of the robot through the sliding bar and show the first person view (FPV) of the robot in the window.

\subsection{Tagging GPR Measurement with Pose Information}
It is very crucial that the GPR data is tagged with robot pose at each GPR sampling that will eliminate the constraint of needing GPR data collection along straight lines in X-Y directions. Allowing the robot to scan in arbitrary and irregular trajectories makes the GPR data collection much easier and facilitates the 3D GPR imaging data analysis.

\begin{figure}[H]
    \centering
    \includegraphics[width=0.5\textwidth]{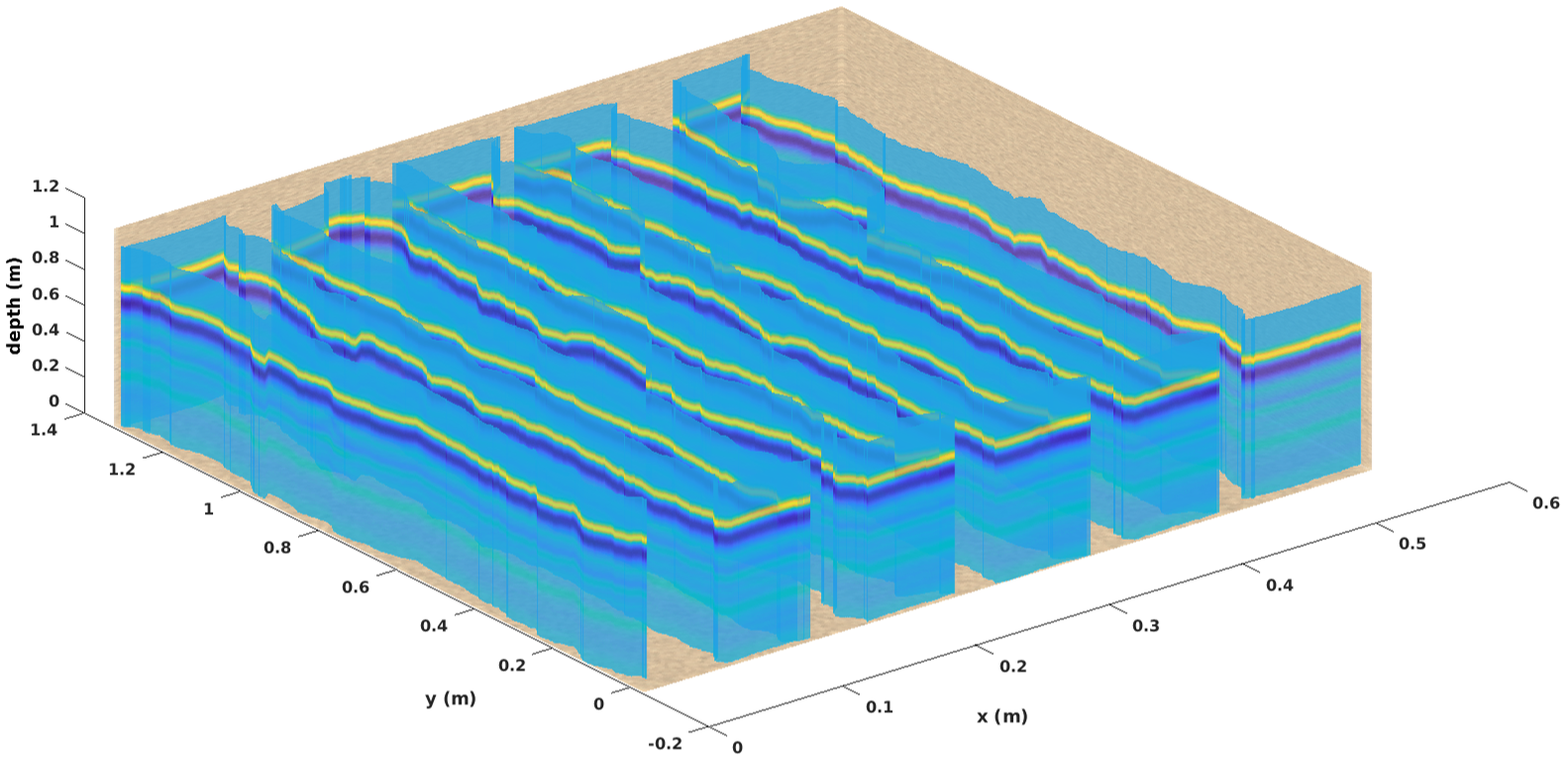}
    \caption{The B-scan profile tagged with metric positioning information when a robot moving along a zig-zag trajectory.}
    \label{fig:metric}
\end{figure}

Specifically, our robot carries an RGB-D camera embedded with an IMU sensor to collect RGB and depth images of the construction surface, together with the corresponding IMU data, e.g., quaternion, angular velocity, and linear velocity. Then, by taking advantage of ORB-SLAM3 \cite{campos2021orb} algorithm, it takes RGB images, depth images as the input to conduct visual odometry and fuses with IMU measurement to perform real-time localization. Then, we implement a time synchronizer function in Robot Operation System (ROS), which takes in messages of different types from multiple sources, and outputs them only if it has received a message on each of those sources with the same timestamp. It is used to synchronize the GPR sampling with vision-based positioning data so that the GPR data collection would not be constrained to the straight line.

The frame rate of the RGB-D camera is 30Hz, and the IMU update rate is 200MHz. Through the interpolation, we achieved 200Hz for position updates. Since the PaveScan GPR sampling rate is 100Hz, we synchronized the vision-based positioning and GPR updating at 100Hz in the experiments. It demonstrates that the vision-based accurate positioning solution has met the “low latency” requirement since 100Hz$-$200Hz is more than good enough for almost all commercial GPR applications. Figure~\ref{fig:metric} illustrates an example of how GPR B-scan data is collected and tagged with pose information in a zig-zag pattern, and it does not require the intervention of the human inspector.

\begin{figure}
    \centering
    \includegraphics[width=0.4\textwidth]{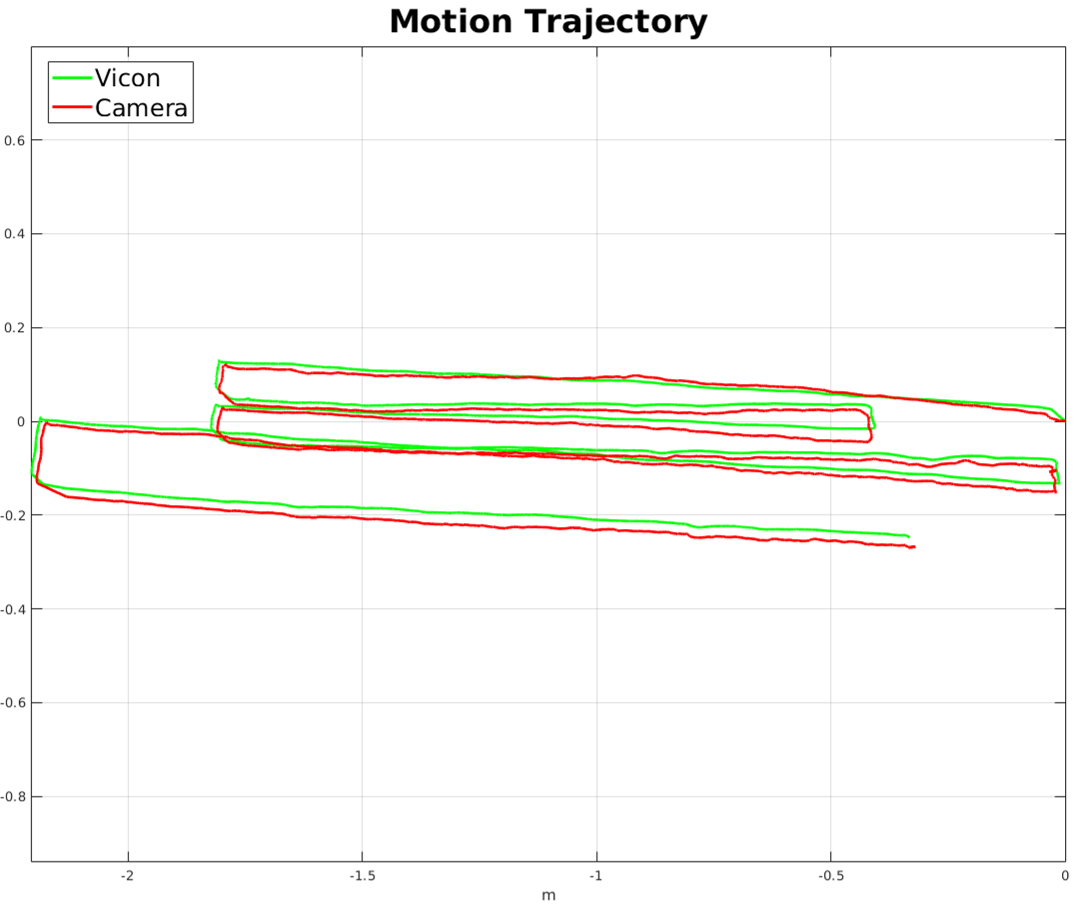}
    \caption{The motion trajectory between ground truth, which is provided by VICON system, and camera. Note that this motion is in a random zig-zag pattern.}
    \label{fig:traj_pos}
\end{figure}

\begin{figure}
    \centering
    \includegraphics[width=0.45\textwidth]{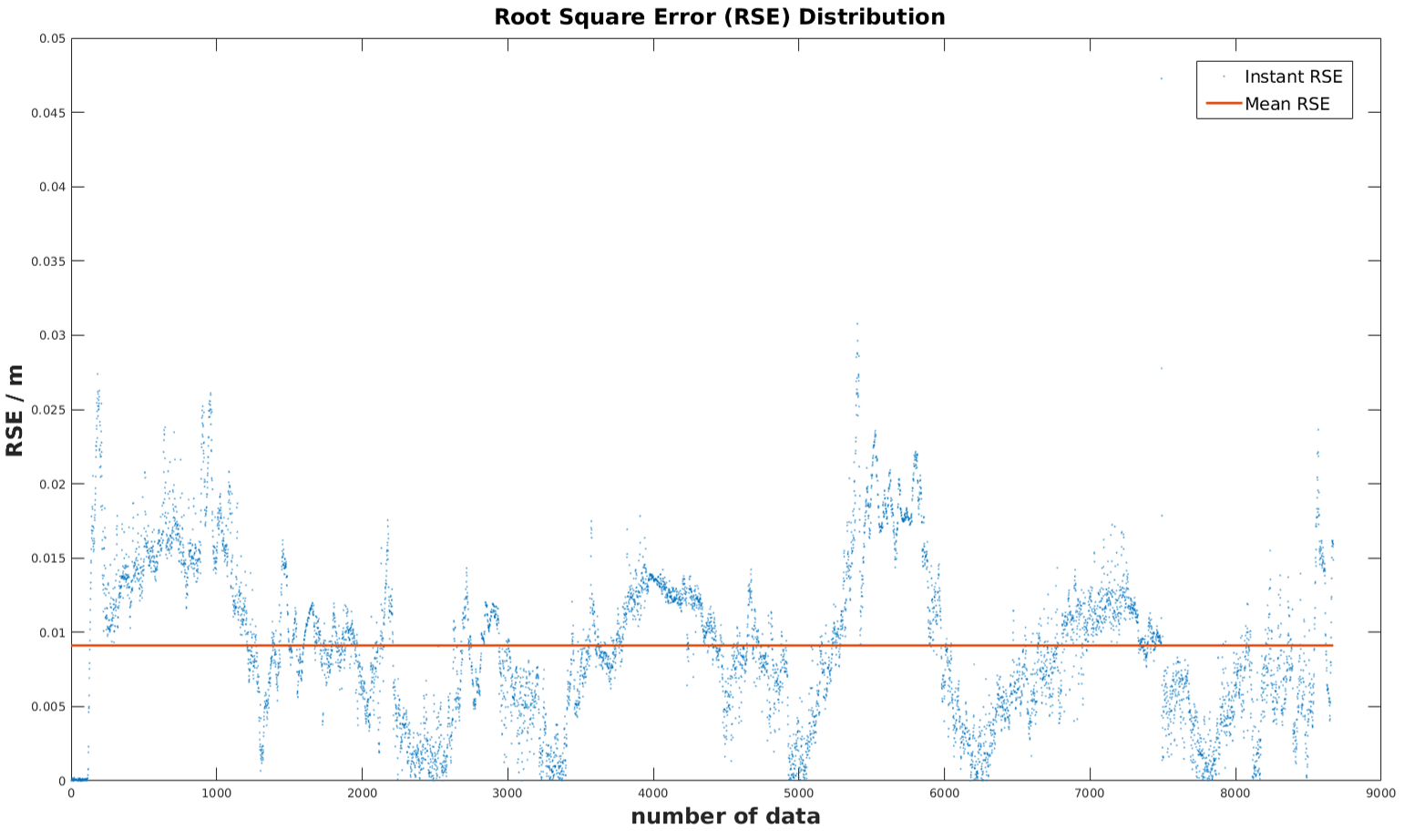}
    \caption{The RSE error distribution between the ground truth motion and estimated motion provided by visual positioning. It shows that the mean RSE error is only 0.9 cm, which could meet the positioning accuracy in practical GPR collection.}
    \label{fig:error_pos}
\end{figure}

Furthermore, we conduct an accuracy test of the robot motion using VICON system \cite{merriaux2017study}. As shown in Figure~\ref{fig:traj_pos}, we control the robot move in a zig-zag pattern, where the green line indicates the ground truth of the motion trajectory provided by the VICON system, and the red line represents the motion trajectory estimated by an RGB-D camera. Figure~\ref{fig:error_pos} and Equ.~\ref{fig:rmse} denote the Root Square Error (RSE) between the ground truth and vision-based trajectory. It can be noted that the mean RSE is less than 1 centimeter, which satisfies the requirement for high accurate positioning by GPR industry.  

\begin{equation}
Mean \ R S E=\sqrt{\frac{\sum_{i=1}^{N}\|y(i)-\hat{y}(i)\|^{2}}{N}}
\label{fig:rmse}
\end{equation}

Note that $N$ indicates the number of position samples while $i$ means the $i$-th position sample. $y(i)$ and $\hat{y}(i)$ represent the ground truth position and estimated position respectively.

\begin{figure*}
    \centering
    \includegraphics[width=0.9\textwidth]{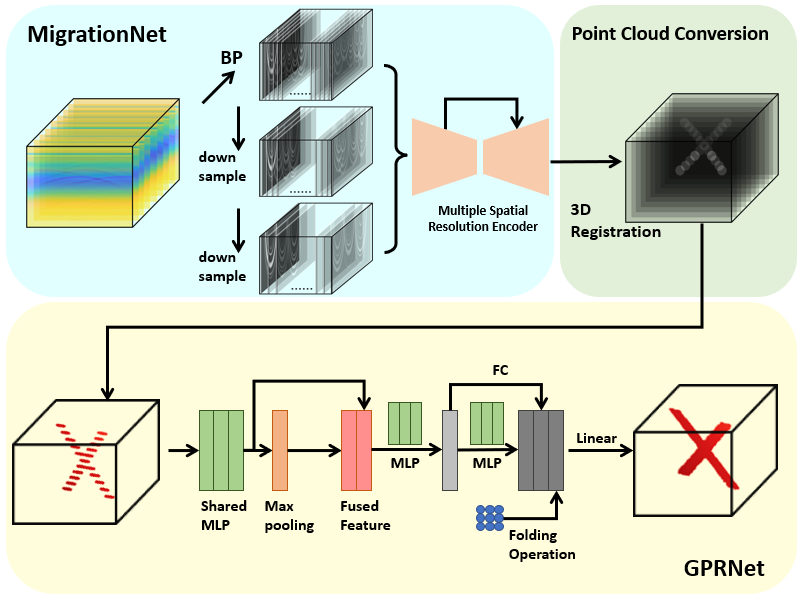}
    \caption{Network architecture. MigrationNet aggregates the BP data with multiple resolutions, and interprets those BP data into a set of 2D images that indicate the cross-section of the subsurface utilities. Then, the cross-section images are converted into the sparse point cloud while the GPRNet completes the sparse point cloud to make it dense and continuous.}
    \label{fig:networks}
\end{figure*}

\section{MigrationNet for GPR Data Interpretation}
\label{section:migrationnet}
We present the learning-based GPR data processing methods as shown in Figure~\ref{fig:networks}, which consists of MigrationNet, GPRNet, as well as a point cloud conversion process. In this section, we train a network, MigrationNet, to interpret the input B-scan data to a cross-section image of the underground object model. 

Given the $k$-th B-scan data $\mathcal{B}^{k}=\left\{\left(\mathcal{A}_{i}^k, \mathbf{T}_{i}\right)\right\}_{i=1}^{N} \in \mathbb{R}^{N \times M \times 1}$ where $k=1,...,K$, we assume $\mathcal{B}^{k}$ consists of $N$ tuples of A-scans $\mathcal{A}_{i}^k =\left\{\left(\mathbf{a}_{j}, \mathbf{t}_{j}\right)\right\}_{j=1}^{M} \in \mathbb{R}^{M \times 1}$ along with their corresponding pose $\mathbf{T}_{i} \in \mathbb{R}^{3 \times 4}$ (where in $\mathcal{A}_{i}^k$, $M$ indicates the number of samples in an A-scan data; additionally, $\mathbf{a}_{j}$ and $\mathbf{t}_{j}$ denotes the amplitude and the traveling time of the $j$-th A-scan sample). Our goal is to distill B-scan data into a BP-based representation $\Theta$. In addition, we are interested in the GPR data processing $\Theta$ that allows us to interpret the $\Theta({\mathcal{B_{k}}})$ into a clear, user-friendly cross-section image of the underground object model. In the following, we first formalize $\Theta$ and then discuss the details of interpretation algorithm in $\Phi$. In all, Algorithm~\ref{alg:Framwork} describes the processing of this approach, which is called MigrationNet. 

\renewcommand{\algorithmicrequire}{ \textbf{Input:}} 
\renewcommand{\algorithmicensure}{ \textbf{Output:}} 
\begin{algorithm}[htb] 
\caption{ MigrationNet For GPR Data Interpretation \cite{feng2021subsurface}} 
\label{alg:Framwork} 
\begin{algorithmic}[1] 
\REQUIRE ~~\\ 
B-scan data $\mathcal{B}^{k}=\left\{\left(\mathcal{A}_{i}^k, \mathbf{T}_{i}\right)\right\}_{i=1}^{N} \in \mathbb{R}^{N \times M \times 1}$;\\
\ENSURE ~~\\ 
A cross-section image $\mathcal{I}^k$ of the subsurface utility's geometry;
\STATE for k = 1, 2, ..., n
\STATE \ \ Downsample the B-scan data $\mathcal{B}^{k}$ to $\mathcal{\hat{B}}^{k}$
\label{ code:fram:extract }
\STATE \ \ Sparse back-project $\mathcal{\hat{B}}^{k}$ via function $\Theta$ to get the input $\mathbf{Z}^{k}$.
\label{code:fram:trainbase}

\STATE \ \ Extract $\left\{{\mathbf{f}}_{i}\right\}_{i=1}^{N=3} \in \mathbb{R}^{M \times N \times 512}$ from $\left\{\mathcal{E}_{i}^M\right\}_{i=1}^{N=3}$; 
\STATE \ \ Estimate $\mathcal{I}^k$ through the decoder.
\STATE end
\RETURN $\Phi(\mathbf{Z}^{k})=\mathcal{I}^k$; 
\end{algorithmic}
\end{algorithm}

\subsection{Sparse Back Projection Aggregation}
As discussed in Section~\ref{section:related_works}, BP approach $\Theta$ serves as a common solution to process the GPR data\cite{li2019toward}. However, there are some limitations existing in BP approach. First of all, it needs to back-project each A-scan $\mathcal{A}_{i}^k$ into a pre-defined 3D volumetric map. Second, each A-scan $\mathcal{A}_{i}^k$ covers a cone-volume of subsurface, and it usually overlaps with other neighbor A-scans. The back-projection of each A-scan would result in heavy computation during the fusion because of the indexing, which is computation expensive and requires large memory to support the computation. 

In order to address this challenge, we introduce a multi-spatial resolution algorithm to aggregate the A-scans, where the resolution denotes how many A-scan measurements $\hat{N}$ ($0 < \hat{N} < N$) from a $\mathcal{B}^{k}$ are used for back-projection. In particular, since each $\mathcal{B}^{k}$ might have a different number of A-scan measurements $N$, for any $\mathcal{B}^{k}$ whose number of A-scan measurements are less than $1,024$, we only pick $\hat{N}$ = $256$, $128$ and $64$ A-scan measurements for back-projection and stack them in the spatial domain as the input. This is how we distinguish the different spatial resolution in the input data. Otherwise, for those $\mathcal{B}^{k}$ with more than $1,024$ A-scan measurements, we introduce a sliding-window crop operation to split the B-scan data into multiple segments. As introduced in Equ.~\ref{equ:trim}, we fix the sliding window length to $1,024$, where $q$ is equal to the ceiling value of this equation, which represents the number of cropped B-scans after the trim operation.
\begin{align}
    q = \lfloor N/1024 \rfloor
    \label{equ:trim}
\end{align}

\begin{figure}[h]
    \centering
    \includegraphics[width=0.3\textwidth]{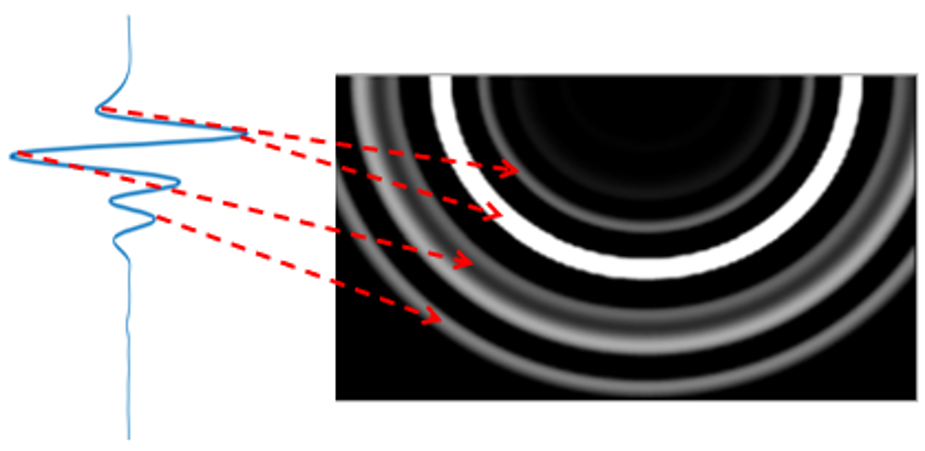}
    \caption{ BP algorithm converts the A-scan raw data into a set of semi-spheres.}
    \label{fig:bp_processing}
\end{figure}

Then, by taking advantage of the BP algorithm, each sample's amplitude in an A-scan is converted into a \emph{semi-sphere} shape at its corresponding traveling time. As illustrated in Figure~\ref{fig:bp_processing}, the brighter semi-sphere indicates the higher amplitude part in A-scan. Furthermore, the radius of each semi-sphere in BP image indicates the depth between the ground and the object, which is illustrated by Equ.~\ref{equ:bp}\cite{li2019toward}: 

\begin{equation}
\begin{array}{r}
\forall \mathcal{A}_{i}^{k} \in \mathcal{\hat{B}}^{k},\ \left(x-\mathbf{t}_{x}\right)^{2}+\left(y-\mathbf{t}_{y}\right)^{2}=\left(\mathbf{a}_{j} * \mathbf{t}_{j}\right)^{2} \\
y<0, \ 1<i<N, \ 1<j<M
\end{array}
\label{equ:bp}
\end{equation}
where $\mathbf{t}_{x}$ and $\mathbf{t}_{y}$ represent the position of the current A-scan $\mathcal{A}_{i}^k$. In particular, given a 2D back-projected data $\mathcal{P}_{i}^{k}$ converted from $\mathcal{A}_{i}^{k}$ as shown in Figure~\ref{fig:bp_processing}, its height and width are the same as a raw B-scan $\mathcal{B}^{k}$, which are equal to $M$ and $N$ respectively. It is because BP is an algorithm that aggregates each back-projected data $\mathcal{P}_{i}^{k}$ from a B-scan $\mathcal{B}^{k}$, and the intersection part in the aggregated BP data $\sum_{i=1}^{N} \mathcal{P}_{i}^{k}$ with the highest energy level indicates a potential target.

In summary, we leverage on the BP principle to represent and process the cropped B-scan $\mathcal{\hat{B}}^{k}$ as a function $\Theta$:

\begin{equation}
\resizebox{.9\hsize}{!}{$\Theta: \mathbb{R}^{\hat{N} \times M \times 1} \rightarrow \mathbb{R}^{\hat{N} \times M \times N}, \quad {\mathcal{\hat{B}}^{k}} \mapsto \Theta({\mathcal{\hat{B}}^{k}})=\sum_{i=1}^{\hat{N}} \mathcal{P}_{i}^{k} = \mathbf{Z}^{k}$}
\end{equation}

In this way, a sparse-stacked multi-resolution input $\mathbf{Z}^{k}$ is created. We choose to sparsely aggregate the back-projected data $\mathcal{P}_{i}^{k}$ because a sparse data can decrease the computational cost, and can provide multiple resolution of the input data in the spatial domain. We provide more details and experiments in Section \ref{section:ablation}. 

\subsection{MigrationNet Formulation}
Given a sparse-stacked input $\mathbf{Z}^{k}$, we introduce a interpretation algorithm $\Phi$, that maps a sparse-stacked B-scan data to an image $\mathcal{I}^k$.

\begin{equation}
\Phi: \mathbb{R}^{\hat{N} \times M \times N} \rightarrow \mathbb{R}^{M \times N \times 1}, \quad\Theta({\mathcal{\hat{B}}^{k}}) \mapsto \Phi(\mathbf{Z}^{k})=\mathcal{I}^k
\end{equation}

In particular, $\Phi$ is composed of an encoder, which has multiple spatial resolution to extract features from multiple resolution input $\mathbf{Z}^{k}$, and a decoder, to interpret the features and predict a cross-section image $\mathcal{I}^k$ of the subsurface utility's geometry. 

\subsubsection{Multiple Spatial Resolution Encoder}
A multi-resolution representation is the key for $\Phi$ to interpret $\mathbf{Z}^{k}$. Thus, we first introduce our feature extractor, named Multiple Spatial Resolution Encoder (MSRE). Here, we take inspiration from the Feature Pyramid Network (FPN)\cite{lin2017feature} structure. FPN belongs to the class of object detection algorithms, which inherits the feature capture ability by introducing a multi-resolution input, and reveals the rich local structure information in the spatial domain. 

Specifically, the input $\mathbf{Z}^{k} \in \mathbb{R}^{\hat{N} \times M \times N}$ has three resolutions where $\hat{N}=\{256, 128, 64\}$. Thus, to extract features from $\mathbf{Z}^{k}$, we use three independent feature extractors $\left\{\mathcal{E}_{i}^M\right\}_{i=1}^{N=3}$ composed of several down-sampling groups $\left\{{\mathbf{d}}_{i}\right\}_{i=1}^{N=3}$. Note that each down-sampling group $\left\{{\mathbf{d}}_{i}\right\}_{i=1}^{N=3}$ is a combination of two convolution layers and one max-pooling layer while the kernel size $\{s_i^k\}_{i=1}^{N=3}$ of the max-pooling layer is different. Specifically, when $\hat{N} = 256$, the kernel size of the max-pooling layer in the corresponding extractor is equal to $s_1^k = 8$. In addition, the kernel size in the max-pooling layer is set to $s_2^k = 4$ when $\hat{N} = 128$ while the max-pooling kernel size is equal to $s_3^k = 2$ when $\hat{N} = 64$. Hence, three corresponding latent feature map $\left\{{\mathbf{f}}_{i}\right\}_{i=1}^{N=3} \in \mathbb{R}^{M \times N \times 512}$ are extracted from $\left\{\mathcal{E}_{i}^M\right\}_{i=1}^{N=3}$.

\subsubsection{Decoder}
To estimate a cross-section image $\mathcal{I}^k$ from the extracted features, here we introduce the decoder frame work. Given our current feature maps $\left\{{\mathbf{f}}_{i}\right\}_{i=1}^{N=3} \in \mathbb{R}^{M \times N \times 512}$, we concatenate them to $\mathbf{F} \in \mathbb{R}^{M \times N \times 1536}$ and pass through the encoder. In particular, this decoder is composed of 4 up-sampling groups $\left\{{\mathbf{u}}_{i}\right\}_{i=1}^{N=4}$, and each group contains two convolutional layers and one deconvolutional layer. Besides, we also take advantage of skip connections, that skip-connects features between the $\left\{{\mathbf{d}}_{i}\right\}_{i=1}^{N=3}$ and $\left\{{\mathbf{u}}_{i}\right\}_{i=1}^{N=4}$.

To summarize, the decoder interprets the concatenated feature map $\mathbf{F}^M \in \mathbb{R}^{M \times N \times 1536}$ to a cross-section binary image $\mathcal{I}^k \in \mathbb{R}^{M \times N \times 1}$, where the white region indicates the cross-section of the utilities and the black area indicates the background.

\subsubsection{Loss Design and Training}
\label{section:loss}
To constrain the shape and size of the underground cylindrical objects, we leverage on a joint loss $\mathcal{L}$ that consists of two terms: a structure similarity loss and a cross entropy loss.

In particular, in most of the real non-destructive testing test cases, objects such as rebars, utilities, and PVC pipes all have a round shape cross-section. Hence, it is necessary to compare structure similarity between the predicted image and the ground truth to maintain the proper size and shape. Inspired by \cite{wang2004image,zhao2015loss,godard2017unsupervised}, we demonstrate the structure comparison loss between predicted image $X$ and ground truth $Y$ as follows:

\begin{equation}
\mathcal{L}_{S} = \frac{\sigma_{xy} + {C}}{\sigma_{x}\sigma_{y} + {C}}
\end{equation}

where $\sigma_{x}$ and $\sigma_{y}$ are the standard deviation as an estimate of the image contrast, $C$ is a constant value while $\sigma_{xy}$ represents the covariance which is:

\begin{equation}
\sigma_{xy} = \frac{1}{M*N-1}\sum_{q=1}^{M*N}(x_{q}-\mu_{x})(y_{q}-\mu_{y})
\end{equation}
Note that  $\mu_{x}$ and $\mu_{y}$ are mean intensity of the predicted image $\mathcal{I}^k$ and ground truth respectively while $x_{q}$ and $y_{q}$ indicates each pixel's coordinate. In addition, $M*N$ indicates the number of pixels in the image.

We then use cross entropy loss \cite{ronneberger2015u} as the second loss expression in this joint loss design:

\begin{equation}
{\mathcal{L}_{CE}} = \sum_{x_{k} \in \mathcal{M}} w^{l}(x)log(p(x_{k,l}))
\label{equ:unetcrossentryLoss}
\end{equation}

where $x_{k}$ indicates an element in given input while ${M}$, $p(x_{k, l})$ is the element $x_{k}$ probabilistic prediction over class $l$, and $w^{l}$ is the weight of each classes.

Finally, our loss function is expressed in Equ.~\ref{equ:loss_function}. $\lambda_{i}$ and $\lambda_{j}$ denote the weight of cross entropy loss and structure loss that satisfy the relation as $\lambda_{i}$ + $\lambda_{j}$ = 1.

\begin{equation}
{\mathcal{L}} = \lambda_{i} \mathcal{L}_{S} + \lambda_{j} \mathcal{L}_{CE}
\label{equ:loss_function}
\end{equation}

\textbf{Training:} The weights governing the terms in loss function is set to $\lambda_{i} = 0.1$ and $\lambda_{j} = 0.9$, we also use the stochastic gradient descent (SGD), select momentum as $0.9$ and weight decay as $1e$-$8$. As for the initial learning rate (LR) and input scale, by evaluating the average accuracy, average precision, average recall as well as F1 score in training dataset with different LR and scale, the learning rate is set to $5e$-$6$ while the input scale is $0.25$.

\section{GPRNet: GPR Pipes Reconstruction Network for 3D Modelling}
\label{section:GPRNET}
In this section, we introduce the algorithm~\ref{alg:Framwork2} which is a 3D modelling network that reconstructs underground pipes (named as GPRNet). 

\renewcommand{\algorithmicrequire}{ \textbf{Input:}} 
\renewcommand{\algorithmicensure}{ \textbf{Output:}} 
\begin{algorithm}[htb] 
\caption{ GPR-based Subsurface Pipe Reconstruction.} 
\label{alg:Framwork2} 
\begin{algorithmic}[1] 
\REQUIRE ~~\\ 
The interpreted cross-section image, $\mathcal{I}^{k} \in \mathbb{R}^{M \times N \times 1}$;\\
The pose associated with B-scan data, $\mathbf{T}^k$;\\
\ENSURE ~~\\ 
Dense point cloud set $\mathbf{P^{D}}$ of the subsurface pipes;
\STATE for k = 1, 2, ..., n
\STATE \ \ Convert the given $\mathcal{I}^{k}$ and its corresponding pose $\mathbf{T}^k$ to a sparse point cloud set $\mathbf{P}$.
\label{ code:fram:extract }
\STATE end
\label{code:fram:trainbase}
\STATE Extract the point feature vector $\left\{\mathbf{v}_{i}\right\}_{i=1}^{N=3}$ from $\left\{\mathcal{E}_{i}^G\right\}_{i=1}^{N=3}$; 
\STATE Estimate $\mathbf{P^{D}}$ through the decoder.
\RETURN $\Psi\left(\mathbf{P}\right)=\mathbf{P^{D}}$; 
\end{algorithmic}
\end{algorithm}

\subsection{From 2D Image to 3D Point Cloud}
\label{subsection:pc_register}

We introduced the GPR interpretation algorithm in last section, we now expect to reveal the spatial information of the subsurface objects' structure based on the interpretation results. Hence, we first register the predicted binary cross-section image set $\mathcal{I}=\{\mathcal{I}^{k} \in \mathbb{R}^{M \times N \times 1}|k=1, 2, ... ,K\}$ into the 3D space according to its pose $\mathbf{T}$, where $\mathbf{T}=\{\mathbf{T}^{k}=\{\mathbf{T}_{i}\}_{i=1}^{N}|k=1, 2, ... ,K\}$ and $k$ represents the $k$-th image/pose corresponding to B-scan data $\mathcal{B}^{k}$. Note that the pose $\mathbf{T}_{i}$ is obtained from the vision-based positioning introduced in Section~\ref{section:data_collection}. In this way, we can make sure for any interpreted image $\mathcal{I}^{k}$, it shares the same pose information $\mathbf{T}^{k}$ as its corresponding B-scan data $\mathcal{B}^{k}$. 

We further convert the registered image set to get a sparse point cloud set: $\mathbf{P}=\left\{p_{i}\right\}_{i=1}^{C}$. We regard this point cloud $\mathbf{P}$ consisting of $C$ points while each point $p_{i} \in \mathbb{R}^{3}$.

Specifically, $\mathcal{I}^{k}$ is a binary image, and the white pixel value is equal to $1$. Thus, we register these 2D white pixels into a 3D space by aggregating multiple images, and the third dimension is provided by visual positioning information that tagged with the image.


Once obtain $\mathbf{P}$, we use iterative farthest point sampling (IFPS), which is a sampling strategy applied in Pointnet++ \cite{qi2017pointnet++} to get a set of skeleton points. IFPS can represent the distribution of the entire point sets better compared to random sampling, and it is more efficient than CNNs\cite{huang2020pf}. After the implementation of IFPS, we evenly distribute each point cloud $\mathbf{P}=\left\{p_{i}\right\}_{i=1}^{C}$ as input to GPRNet, where $C$ equals $1500$.

\subsection{GPRNet Formulation}
Formally, given a set of sparse point cloud $\mathbf{P}=\left\{p_{i}\right\}_{i=1}^{C}$, GPRNet aims to complete the gap between sparse point clouds, and predict a continuous, dense point cloud representation $\mathbf{P^{D}}=\left\{p^{d}_{i}\right\}_{i=1}^{C\prime}$, where $p^{d}_{i} \in \mathbb{R}^{3}$ and $C\prime > C$. We now reason about this procedure through $\Psi$, and $\Psi$ defines a learning-model which has an \textit{encoder decoder structure}. This procedure is illustrated in the following:

\begin{equation}
\Psi: \mathbb{R}^{C \times 3} \rightarrow\mathbb{R}^{C\prime \times 3}, \quad \mathbf{P} \mapsto \Psi\left(\mathbf{P}\right)=\mathbf{P^{D}}
\end{equation}

\subsubsection{Encoder}
Similar to our MigrationNet, GPRNet's encoder $\left\{\mathcal{E}_{i}^G\right\}_{i=1}^{N=3}$ takes advantage of multi-resolution structure. Each subnet $\mathcal{E}_{i}^G$ is a PointNet layer \cite{qi2017pointnet} that consists of three convolutional multi-layer perceptron (MLP) layers and one max-pooling layer. Given the input point cloud set $\mathbf{P}$, each subnet $\mathcal{E}_{i}^G$ extracts the point feature vector $\left\{\mathbf{v}_{i}\right\}_{i=1}^{N=3} \in \mathbb{R}^{C \times j}$, where $j=\{256, 128, 64\}$ respectively.

Then, a max-pooling layer is conducted on each $\left\{\mathbf{v}_{i}\right\}_{i=1}^{N=3}$ to obtain three intermediate features $\left\{\mathbf{g}_{i}\right\}_{i=1}^{N=3}$ with the same dimension as $\left\{\mathbf{v}_{i}\right\}_{i=1}^{N=3}$. Furthermore, we concatenate each point feature vector $\left\{\mathbf{v}_{i}\right\}_{i=1}^{N=3}$ and each intermediate feature $\left\{\mathbf{g}_{i}\right\}_{i=1}^{N=3}$ together and express as a fused feature matrix. At last, another MLP layer extracts the feature map as a $\mathbf{F}^G \in \mathbb{R}^{C \times 896}$, which is represented in $\textit{light grey}$ color as shown in Figure~\ref{fig:networks}.

\subsubsection{Decoder}
The fully-connected decoder \cite{achlioptas2018learning} is good at predicting the global geometry of point cloud but ignores the local features. In contrast to it, FoldingNet decoder \cite{yang2018foldingnet} is good at generating a smooth local feature. Hence, we take advantage of the decoders mentioned above and introduce a decoder with a hierarchical structure similar to \cite{huang2020pf}, which contains the fully-connected (FC) layer, MLP layer and FoldingNet layer. In particular, $\mathbf{F}^G$ is passed through an FC layer as well as an MLP layer, and concatenate together as $\mathbb{F} \in \mathbb{R}^{896 \times 3}$. In addition, we leverage the folding operation, where a patch of $9$ points is generated at each xyz map in the feature $\mathbb{F}$. Thus, we can obtain the detailed output consisting of $896*9$ ($8064$) points. That is to say, a dense point cloud output, $\mathbf{P^{D}}$, is thus generated from our multi-resolution decoder via the fully-connected and folding operations. Note that $\mathbf{P^{D}}=\left\{p^{d}_{i}\right\}_{i=1}^{C\prime}$, $C\prime$ is equal to $8064$.

\subsubsection{Loss Design and Training}

To constrain and compare the difference between the predict point cloud set $\mathbf{P^{D}}$ and the ground truth point cloud set $\mathbf{P^{GT}}$, an ideal loss must be differentiable concerning point locations and invariant to the permutation of the point cloud. In this paper, we use Chamfer Distance (CD) \cite{fan2017point} loss $\mathcal{L}_{CD}$ to calculate the average closest point distance between $\mathbf{P^{D}}$ and $\mathbf{P^{GT}}$, which is shown in Equ~\ref{equ:loss}. Note that $p_{m}$ and $p_{n}$ denotes a point in $\mathbf{P^{D}}$ and $\mathbf{P^{GT}}$ respectively.

\begin{equation}
\resizebox{.9\hsize}{!}{$\mathcal{L}_{CD} = \frac{1}{\mathbf{P^{D}}} \sum_{p_{m} \in \mathbf{P^{D}}} \min \limits_{p_{n} \in \mathbf{P^{GT}}}\|p_{m}-p_{n}\|_{2} + \frac{1}{\mathbf{P^{GT}}} \sum_{p_{n} \in \mathbf{P^{GT}}} \min \limits_{p_{m} \in \mathbf{P^{D}}}\|p_{n}-p_{m}\|_{2}$}
\label{equ:loss}
\end{equation}

The Chamfer Distance finds the nearest neighbor in the ground truth point set. Thus it can force output point clouds to lie close to the ground truth and be piecewise smooth.
\section{Experimental Study}
\label{section:experiment}
In this section, we first introduce the preparation of the field and synthetic GPR data which are used for training and testing. Then we present several experiments such as comparison study, ablation study to demonstrate the effectiveness of our proposed learning-based method.



\subsection{Data Preparation}
\label{subsection:dataset}
To verify the proposed DNN models in this paper, we prepare a GPR B-scan dataset for training and testing purposes. The dataset we provide contains both synthetic and field B-scan data.

\begin{figure}[h]
    \centering
    \includegraphics[width=0.45\textwidth]{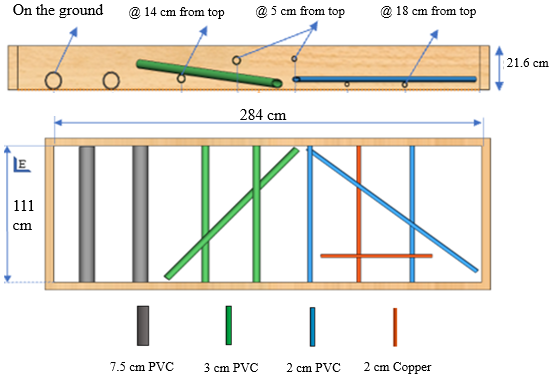}
    \caption{The design details and ground truth of the concrete slab.}
    \label{fig:concrete_slab}
\end{figure}

\subsubsection{Field GPR Data Generation}
We firstly collect the field GPR data with our robotics GPR inspection system on a concrete slab at CCNY Robotics Lab Testing Pit. As mentioned in Section~\ref{section:data_collection}, the GPR sensor we used is a GSSI PaveScan RDM 1.0, with 2 GHz frequency and $20$cm max depth detection range. Figure~ \ref{fig:concrete_slab} shows the design and layout of concrete slab, whose dimension is $2.84(m) \times 1.11(m) \times 0.22(m)$ ($length \times width \times thickness$), and there are 10 pipes embedded in the concrete slab with different size, depth and material. Specifically, 7 pipes are buried horizontally, 1 copper pipe is perpendicular to the other horizontal pipes, and two pipes are diagonally embedded.

We control the Omni-directional robot move along zig-zag path to scan the slab. In the end, there are 24 automated GPR tests are conducted, which contribute 120 field B-scan data to our dataset.

\subsubsection{gprMax Data Generation}
However, the collected field data is still not enough for the DNN model training purpose. Thus, by taking advantage of gprMax\cite{warren2016gprmax}, we build a synthetic testing environment that simulates the real NDT condition. 
Most common assets that buried underground are cylindrical with a round cross-section, for example, rebars, utilities, and PVC pipes. Our simulated environment emulates this condition and involves cylindrical objects with different location, size and direction. Furthermore, in order to match the data collection in field GPR test, we also use a synthetic GPR antenna with 2GHz frequency. At last, we make the spacing of consecutive measurements to 5 mm to match the same property in our field data collection.

\begin{figure}[h]
    \centering
    \includegraphics[width=0.48\textwidth]{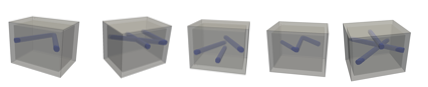}
    \caption{CAD models built by gprMax. All the models emulate the concrete slab property where multiple pipes with different size, direction and depth are inserted in.}
    \label{fig:bscan-intro}
\end{figure}

Specifically, we build $807$ different synthetic concrete slabs with $7263$ B-scan data in gprMax. The simulated GPR pulse is a Gaussian norm wave that has a central frequency $f{_c} = 2G Hz$. The distance between transmitter and receiver of the antenna is set to $5cm$, with a sensing time window as $5 ns$. The surrounding medium of all the concrete slab models is set to a similar value that could mimic the concrete environment, where the relative permittivity is equal to $7$, conductivity is set to $0.01$. Assuming the non-magnetic property of the surrounding environment, thus the relative permeability is set to $1$. Note that all the simulated objects are designed as the Perfect Electric Conductor (PEC). At last, we make the spacing of consecutive measurements to 5 mm to match with the same property in our field data collection. Note that all the slabs have the same dimension, which is: $0.35(m) \times 0.25(m) \times 0.25(m)$ ($length \times width \times thickness$). There are two to six PEC circular-section reinforcing bars buired in each slab with different radius, direction and depth. The above properties make our generalized B-scan dataset have a similar configuration compared to the real GPR data. The front view figure of our synthetic slabs is shown in Figure~\ref{fig:bscan-intro}.

\subsubsection{Ground Truth of Point Cloud Model Generation}
Due to the well-designed field and synthetic slab, we are able to easily generate the ground truth point cloud for training purpose. In particular, since we know the physical property, layout and dimension of each pipe, we can simply calculate the linear equation based on these information. Then, for each point along a line, we adopt a cross-section region within a circle or ellipse. Specially, if an utility pipe is parallel with $x$ or $y$ direction, then the adopted region is a circle, and the radius of this circle $r_c$ is equal to pipe's radius $r$; otherwise, if the pipe is diagonal inserted in a slab with a slope $s$, the cross-section region will be an ellipse, its semi-minor axis length $r_b$ is equal to pipe's radius $r$ while the semi-major axis length $r_a$ is equal to $a*\arctan{s}$.

Note that all points are normalized to have zero-mean per axis and unit-variance. Following prior convention, we generate $8096$ points in each ground truth point cloud set during both training and testing.

\subsection{Experiments Study of MigrationNet}
\subsubsection{Effectiveness of MigrationNet}
\begin{figure*}
    \centering
    \subfigure[Field Raw GPR B-scan image]{
        \includegraphics[width=0.45\textwidth]{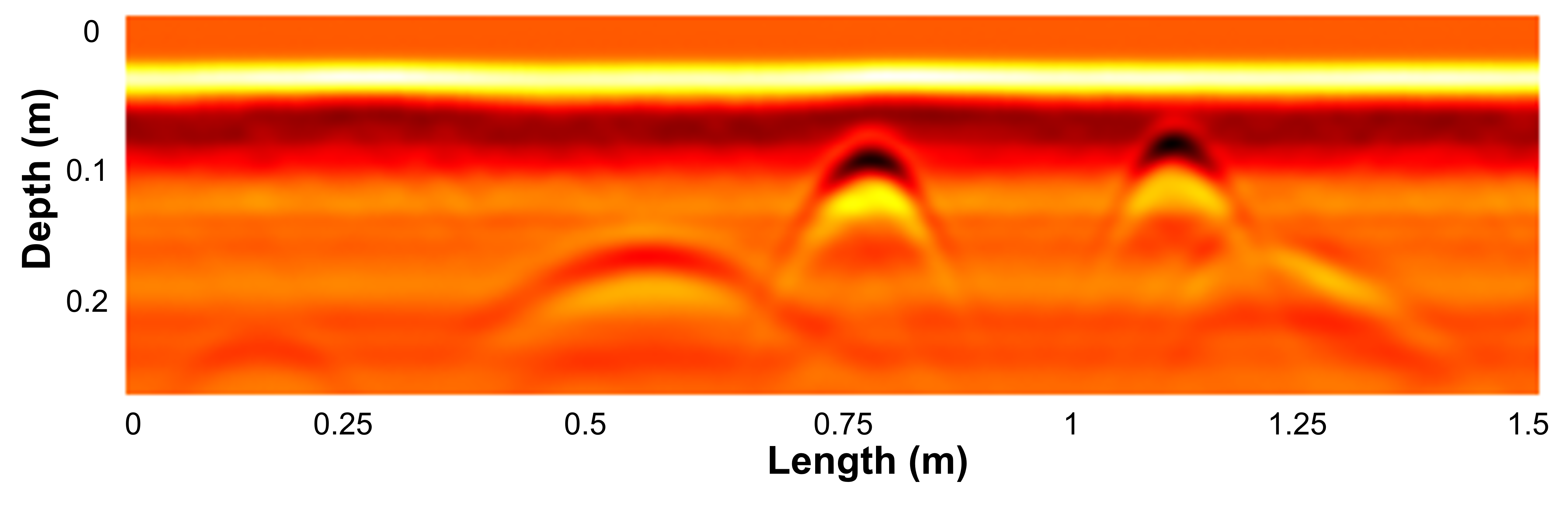}
    }
    \label{fig:exp_c}
    \subfigure[Sparse-BP image aggregated in time domain]{
        \includegraphics[width=0.48\textwidth]{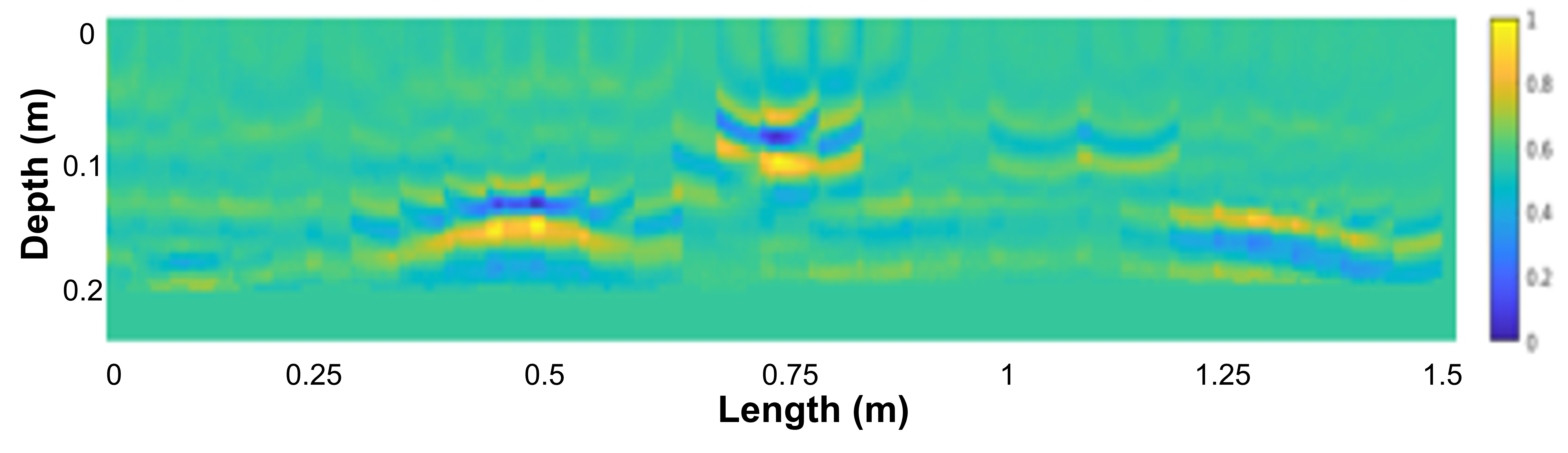}
    }
    \label{fig:exp_a}
    \quad

   \subfigure[Predicted migration result using MirgationNet.]{
        \includegraphics[width=0.45\textwidth]{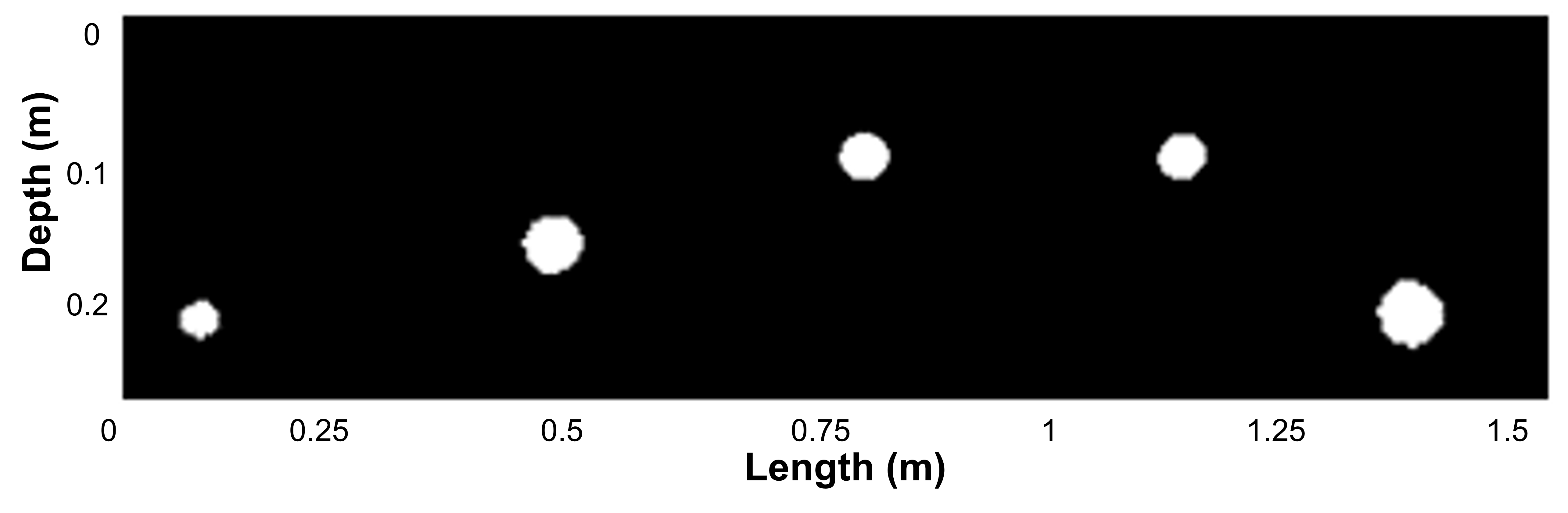}
    }
    \label{fig:exp_e}
    \subfigure[Full-BP image aggregated in time domain]{
    	\includegraphics[width=0.48\textwidth]{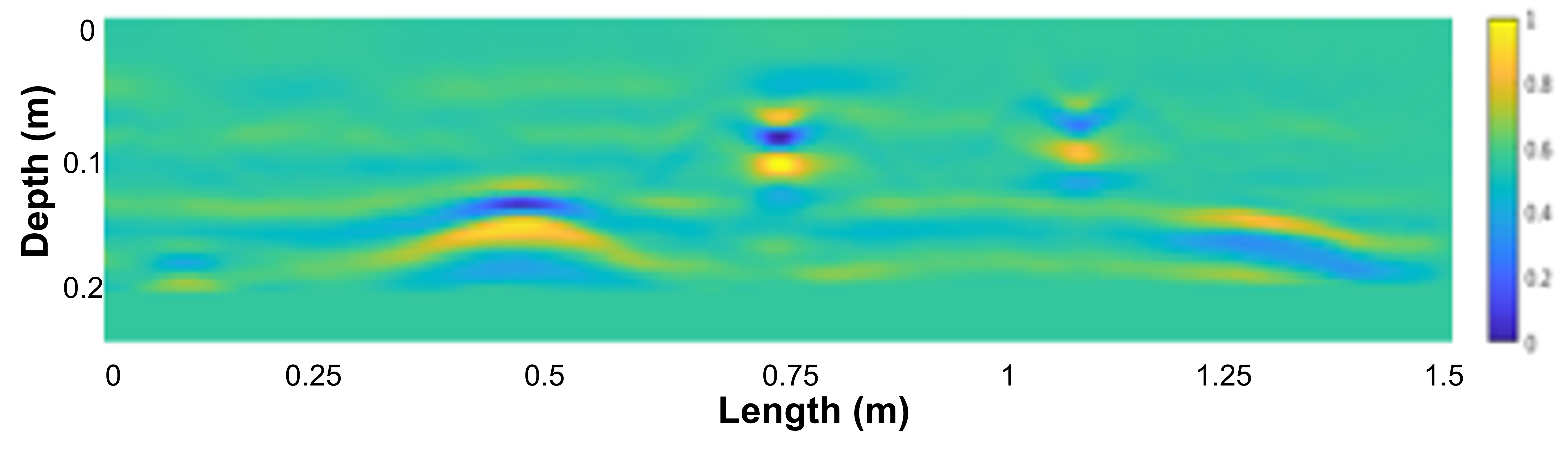}
    }
    \label{fig:exp_b}
    \quad

    \subfigure[Ground-Truth of cross-section image corresponding to its field GPR B-scan data]{
        \includegraphics[width=0.45\textwidth]{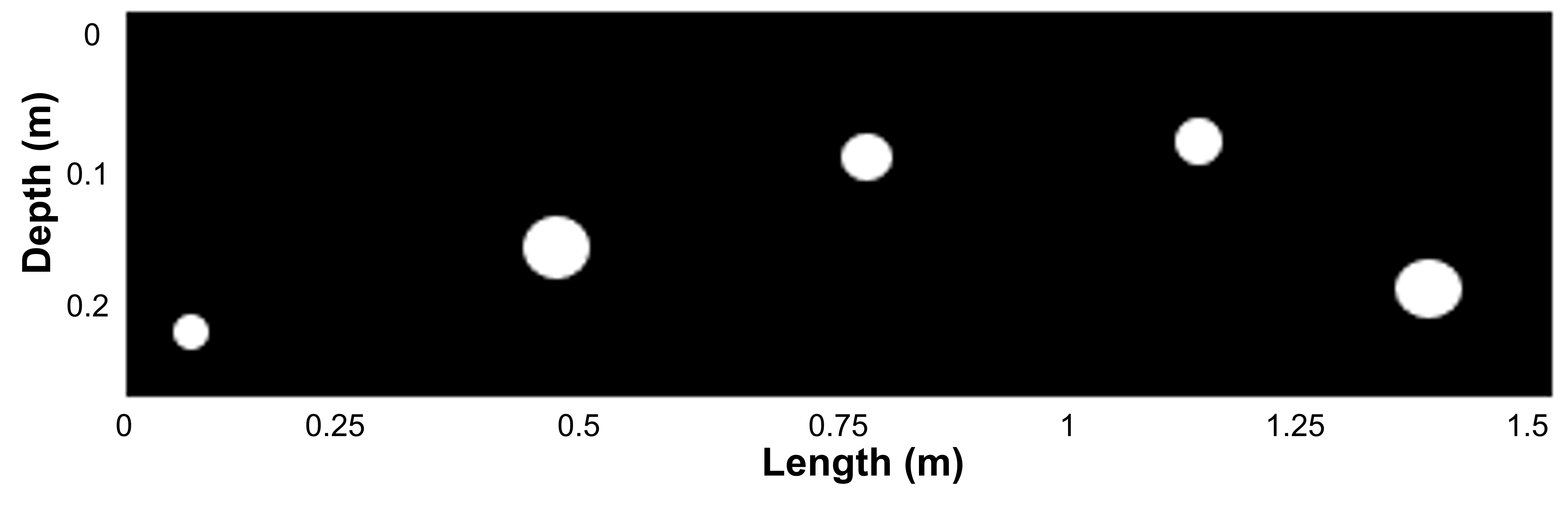}
    }
    \label{fig:exp_d}
    \subfigure[Filtered BP image after applying \emph{Hilbert Transform}.]{
        \includegraphics[width=0.48\textwidth]{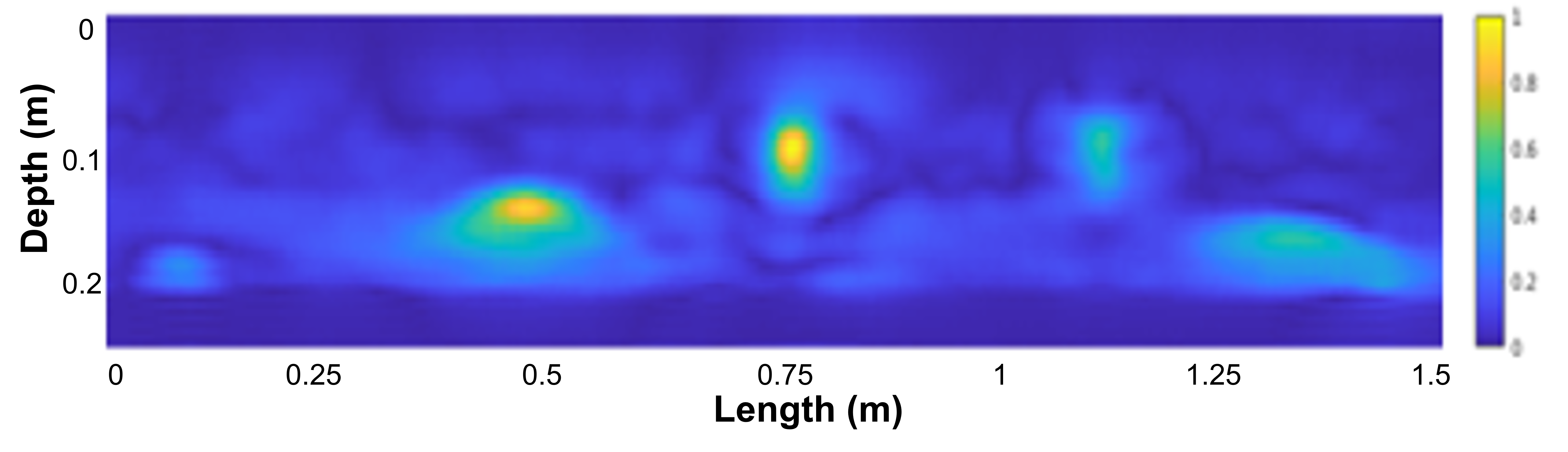}
    }
    \label{fig:exp_f}
    \quad

    \caption{{Qualitative} Migration results comparison between MigrationNet and conventional migration method.}
    \label{Figure.exp}
\end{figure*}

As depicted in Figure~\ref{Figure.exp}, Figure~\ref{Figure.exp} (a) shows the collected on-site GPR B-scan data, which is illustrated in a highlighted \emph{hotmap} format. Figure~\ref{Figure.exp} (b) and (d) represent the back-projected data in the time domain, and are displayed with a highlighted \emph{parula} color code; specifically, Figure~\ref{Figure.exp} (b) shows the conventional migration result using a sparse input BP data, while Figure~\ref{Figure.exp} (d) uses the full BP data as input. We further apply the \emph{Hilbert Transform} filter in Figure~\ref{Figure.exp} (d). The filtered BP image is shown in Figure~\ref{Figure.exp} (f) with a highlighted \emph{parula} color code. Figure~\ref{Figure.exp} (e) indicates the ground truth of cross-section image corresponding to the field B-scan data, while Figure~\ref{Figure.exp} (c) demonstrates the B-scan interpretation result using \textit{MigrationNet}. 

Quantitative effectiveness comparison between the conventional migration and MigrationNet are demonstrated in Table~\ref{table:crnn}. Note that in conventional migration method, the energy level is continuous distributed from 0 to 1. In contrast, the energy level in MigrationNet is binary distributed. That is, 0 stands for the background, and 1 presents the target area. Due to this reason, we convert the conventional migration results to the binary image by selecting the luminance threshold as $0.45$. This luminance threshold would convert the region where energy level is greater than $0.45$ to $1$, and the rest of the region to $0$. In this way, we can compare the GPR image reconstruction results between the conventional and learning-based methods with multiple metrics. 

In particular, we use multiple metrics for the quantitative evaluation. The metrics showed in Equ.~\ref{equ:iou} to Equ.~\ref{equ:ssmi} include \textit{Mean Intersection Over Union (IoU)}, \emph{Pixel Accuracy}, \textit{Mean Square Error (MSE)}, \textit{Signal-to-Noise-Ratio (SNR / dB)} and \textit{Structural Similarity Index (SSMI)}. For metrics \textit{Mean IoU}, \emph{Pixel Accuracy}, \textit{SSMI}, and \emph{MSE}, the larger the value is, the better the performance it stands for; in contrast, for \textit{SNR}, the lower the value is, the better performance it stands for. As shown in Table~\ref{table:crnn}, compared with conventional migration method, MigrationNet gains $30\%$ higher performance in mean IoU, $5.7\%$ higher performance in pixel accuracy, $24.3\%$ higher performance in MSE, $22.2\%$ higher performance in SSMI, and $42.5\%$ less noise in SNR. We can conclude that MigrationNet could effectively improve the performance of GPR imaging reconstruction.

\begin{equation}
\operatorname{loU}\left(S_{t}, S\right)=\frac{S_{t} \cap S}{S_{t} \cup S}=\frac{S_{l} \cap S}{\left|S_{t}\right|+|S|-S_{t} \cap S}
\label{equ:iou}
\end{equation}

\begin{equation}
  \mathrm{Pixel \> Acc.}=\frac{\textit { TP }+\textit { TN }}{\textit { TP }+\textit {TN }+\textit { FP }+\textit { FN }}
\label{equ:acc}
\end{equation}

\begin{equation}
    \mathrm{MSE}=\frac{1}{n} \sum_{i=1, j=1}^{n}\left(X_{i,j}-{Y}_{i,j}\right)^{2}
    \label{equ:mse}
\end{equation}

\begin{equation}
    \mathrm{SNR}=10*(log10(X/Y))
    \label{equ:snr}
\end{equation}

\begin{equation}
    \operatorname{SSIM}(X, Y)=\frac{\left(2 \mu_{X} \mu_{Y}+C_{1}\right)\left(2 \sigma_{X Y}+C_{2}\right)}{\left(\mu_{X}^{2}+\mu_{Y}^{2}+C_{1}\right)\left(\sigma_{X}^{2}+\sigma_{Y}^{2}+C_{2}\right)}
    \label{equ:ssmi}
\end{equation}

Note that in Equ.\ref{equ:iou}, $S_{t}$ is the ground truth region of interests (roi) while $S$ is the predicted roi., In addition, in Equ.\ref{equ:acc}, TP, TN, FP and FN represents \textit{true positive}, \textit{true negative}, \textit{false positive}, \textit{false negative} of the pixel label respectively. Equ.\ref{equ:mse}, $i,j$ denotes pixel index in image $X$ and image $Y$; in Equ.\ref{equ:snr}, we take $X$ as noise signal and compare it r.w.t ground truth image $Y$. In Equ.\ref{equ:ssmi}, $\mu_{X}$, $\mu_{Y}$, $\sigma_{X},$ $\sigma_{Y}$, and $\sigma_{X Y}$ are the local means, standard deviations, and cross-covariance for images $X, Y$, $C_{1}$ and $C_{2}$ are constant values. 

\begin{table}[!th]
\caption{{Quantitative Results on MigrationNet.} Migration Effectiveness Comparison between Conventional Migration and MigrationNet}
\begin{center}
\begin{tabular}{|c|c|c|c|c|c|c|c|c|c|c|}
\hline
\hline
Metrics &\multicolumn{4}{|c|}{Conventional Migration} &\multicolumn{4}{|c|}{MigrationNet} \\ 
\hline
\hline
Mean IoU & \multicolumn{4}{|c|}{62.99}&\multicolumn{4}{|c|}{\textbf{89.97}}\\
\hline
Pixel Acc \% & \multicolumn{4}{|c|}{90.48} &\multicolumn{4}{|c|}{\textbf{95.70}}\\
\hline
MSE & \multicolumn{4}{|c|}{{531.71}} &\multicolumn{4}{|c|}{\textbf{661.13}}\\
\hline
SSMI & \multicolumn{4}{|c|}{0.770} &\multicolumn{4}{|c|}{\textbf{0.941}}\\
\hline
SNR / dB & \multicolumn{4}{|c|}{5.747} &\multicolumn{4}{|c|}{\textbf{3.307}} \\
\hline
\end{tabular}
\end{center}
\label{table:crnn}
\end{table}

\begin{figure*}
    \centering
    \subfigure[Migration result using MigrationNet with $Salt \& Pepper$ noised input, noise variance = $0.05$ ]{
        \includegraphics[width=0.45\textwidth]{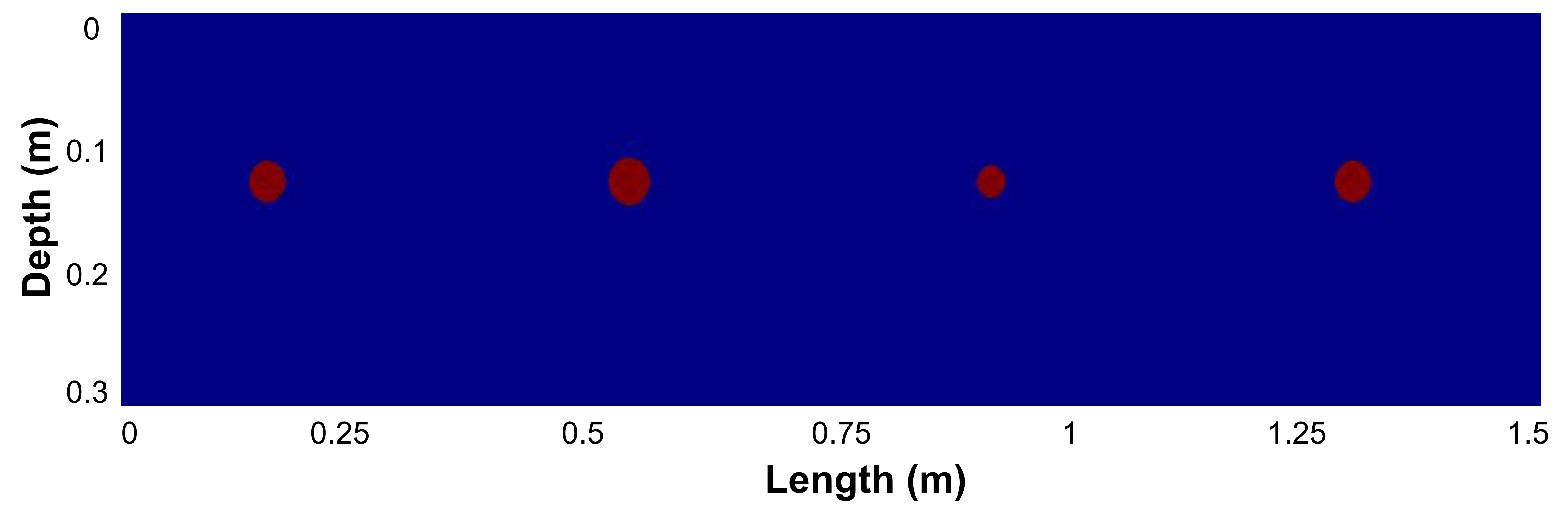}
    }
    \label{fig:fig:noise_pro_pre}
    \subfigure[Migration result using MigrationNet without $Salt \& Pepper$ noised input]{
    	\includegraphics[width=0.45\textwidth]{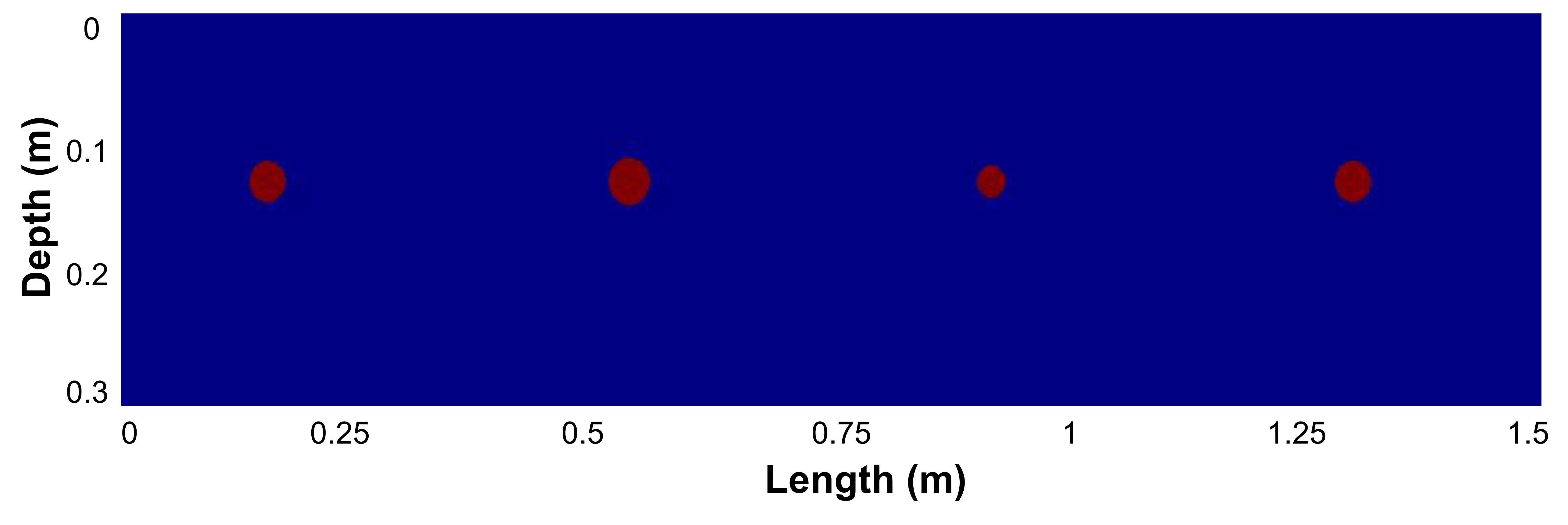}
    
    }
    \label{fig:fig:noise_pro_gt}
    \quad
    
    \subfigure[Conventional Migration result with $Salt \& Pepper$ noised input, noise variance = $0.05$]{
        \includegraphics[width=0.45\textwidth]{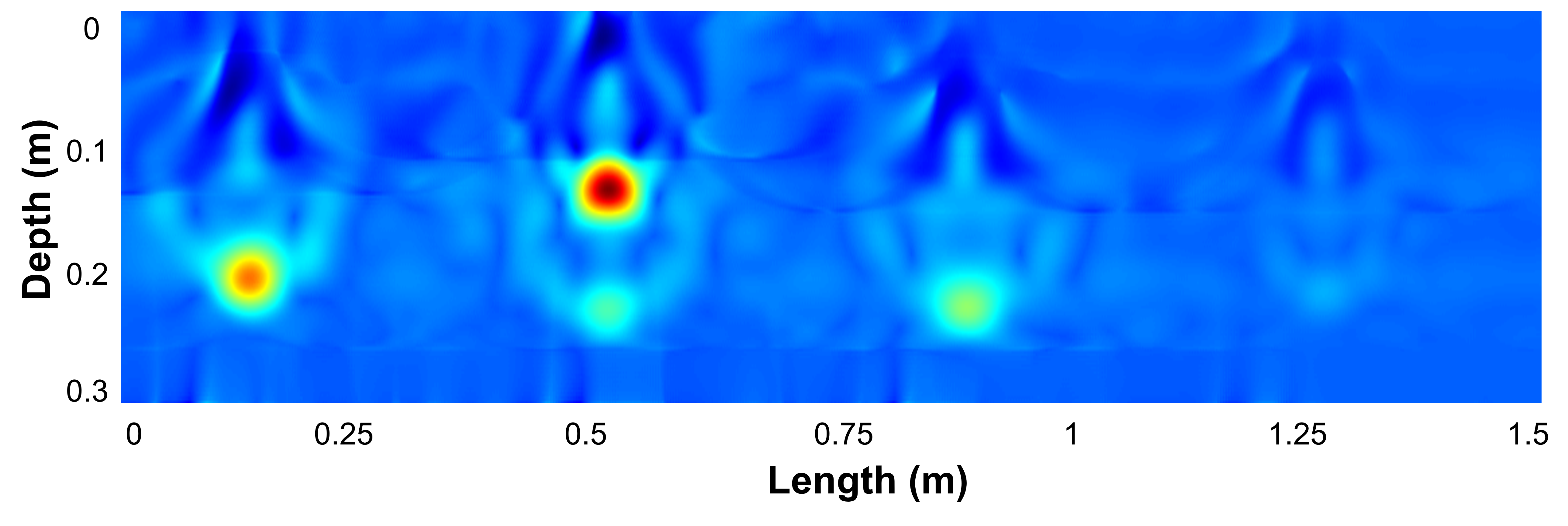}
    }
    \label{fig:fig:noise_tra_pre}
    \subfigure[Conventional Migration result without $Salt \& Pepper$ noised input]{
        \includegraphics[width=0.45\textwidth]{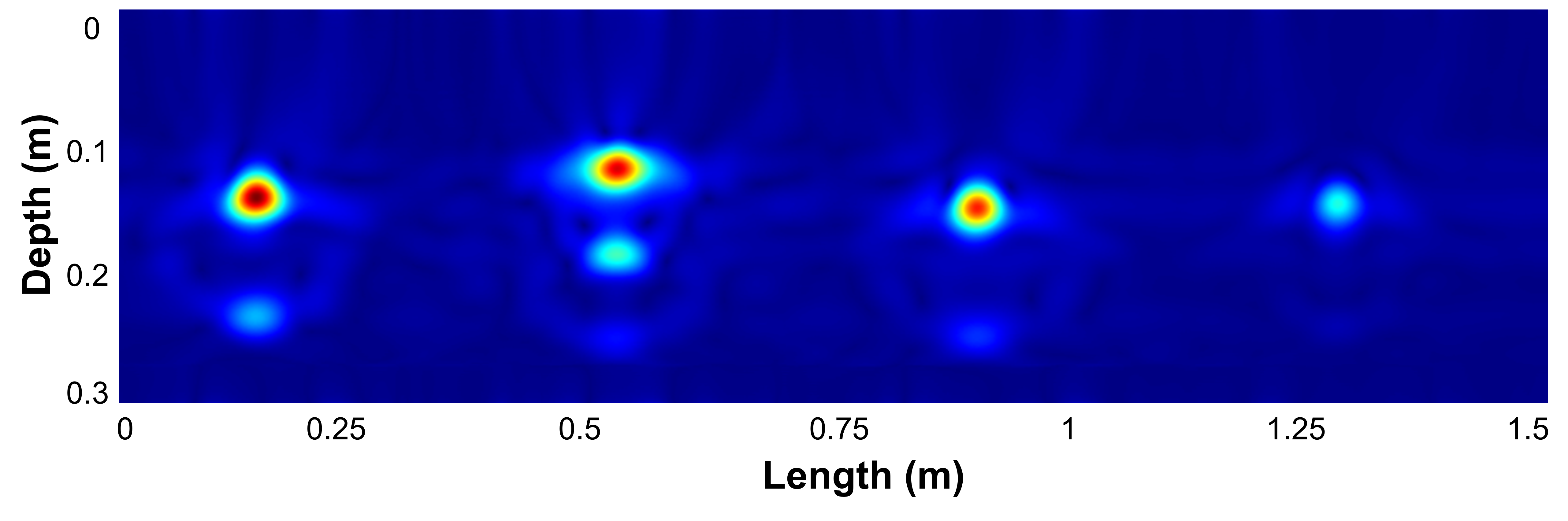}
    }
    \label{fig:noise_tra_gt}
    \quad

    \caption{{Qualitative }Noise Robustness Comparison between conventional and proposed migration method with/without $Salt \& Pepper$ noised input.}
    \label{fig.noise_comparision}
\end{figure*}  

\begin{table*}
\caption{Noise Robustness Comparison. We compare the root-mean-square error (RMSE) between Conventional Migration and MigrationNet.}
\label{table:Migration_NosieComparision}
\begin{center}
\begin{tabular}{|c|c|c|c|c|c|c|c|c|c|c|c|c|c|c|}
\hline
\hline
& \multicolumn{6}{|c|}{Conventional Migration} & \multicolumn{6}{|c|}{MigrationNet } \\ 
\hline
&\multicolumn{2}{|c|}{Gaussian} &\multicolumn{2}{|c|}{Salt \& Pepper} &  \multicolumn{2}{|c|}{Speckle} &\multicolumn{2}{|c|}{Gaussian} &\multicolumn{2}{|c|}{Salt \& Pepper} &  \multicolumn{2}{|c|}{Speckle}
\\ 
\hline
\hline
Without Noise  &  \multicolumn{6}{|c|}{37.3491} & \multicolumn{6}{|c|}{3.3500} \\
\hline
Variance \& Noise density = 0.05  &  \multicolumn{2}{|c|}{54.3589}  & \multicolumn{2}{|c|}{51.6030} & \multicolumn{2}{|c|}{56.1675} & \multicolumn{2}{|c|}{11.4624} & \multicolumn{2}{|c|}{11.2508} & \multicolumn{2}{|c|}{10.2708} \\
\hline
Variance \& Noise density = 0.1 & \multicolumn{2}{|c|}{62.2094} & \multicolumn{2}{|c|}{61.1385} & \multicolumn{2}{|c|}{61.8539}  & \multicolumn{2}{|c|}{{17.8093}}  & \multicolumn{2}{|c|}{16.3628}& \multicolumn{2}{|c|}{16.0731} \\
\hline
Variance \& Noise density = 0.2 &  \multicolumn{2}{|c|}{75.3084} &  \multicolumn{2}{|c|}{77.7894} & \multicolumn{2}{|c|}{76.1743} &  \multicolumn{2}{|c|}{32.1583} & \multicolumn{2}{|c|}{30.9074} & \multicolumn{2}{|c|}{29.5939} \\
\hline
Variance \& Noise density = 0.5 & \multicolumn{2}{|c|}{92.4765} & \multicolumn{2}{|c|}{90.1059} & \multicolumn{2}{|c|}{92.0384}  &  \multicolumn{2}{|c|}{45.3853} & \multicolumn{2}{|c|}{42.8437} & \multicolumn{2}{|c|}{41.2759} \\
\hline
\end{tabular}
\end{center}
\end{table*}

\begin{table}[htbp]
\centering
\caption{Evaluation Performance Comparison with different spatial resolution input of MigrationNet on three metrics.}
\label{table:weight}
\begin{center}
\setlength{\tabcolsep}{1mm}{
\begin{tabular}{|c|c|c|c|cc|c|c|c|c|c|}
\hline
\hline
Multi-Res. Input Channels &\multicolumn{2}{|c|}{MSE} &\multicolumn{2}{|c|}{SSMI} &\multicolumn{2}{|c|}{SNR / dB}\\ 
\hline
\hline
\textbf{256+128+64}    & \multicolumn{2}{|c|}{{661.1313}} & \multicolumn{2}{|c|}{{0.9413}} & \multicolumn{2}{|c|}{\textbf{3.3066}}\\
\hline
256+128   & \multicolumn{2}{|c|}{717.1609} & \multicolumn{2}{|c|}{0.9250} & \multicolumn{2}{|c|}{3.5947}\\
128+64   & \multicolumn{2}{|c|}{832.5777} & \multicolumn{2}{|c|}{0.9199} & \multicolumn{2}{|c|}{5.7474}\\
256    & \multicolumn{2}{|c|}{1.4553e+03} & \multicolumn{2}{|c|}{0.9035} & \multicolumn{2}{|c|}{7.1199}\\
128   & \multicolumn{2}{|c|}{1.433e+03} & \multicolumn{2}{|c|}{0.9075} & \multicolumn{2}{|c|}{7.0553}\\
64   & \multicolumn{2}{|c|}{-} & \multicolumn{2}{|c|}{-} & \multicolumn{2}{|c|}{-}\\
raw input  & \multicolumn{2}{|c|}{\textbf{630.7042}} & \multicolumn{2}{|c|}{\textbf{0.9565}} & \multicolumn{2}{|c|}{3.3849}\\
\hline
\end{tabular}
}
\end{center}
\label{Table:channel_comparison}
\end{table}

We then provide the noise robustness test of MigrationNet. To do this, we add \emph{Gaussian white noise}, \emph{salt \& pepper noise} and \emph{speckle noise} respectively to the input GPR data. Each type of the noise has 4 different variance and noise density parameters, which are $0.05$, $0.1$, $0.2$, and $0.5$. We finally compared \emph{root-mean-square error (RMSE)} metric on conventional migration method and MigrationNet. As illustrated in Table~\ref{table:Migration_NosieComparision} and Figure~\ref{fig.noise_comparision}, we could find that our proposed method has high noise robustness. In contrast, the noise would significantly degrade the migration results when deploying the conventional method,. 

\subsubsection{Ablation Study for MigrationNet}

\label{section:ablation}
\textbf{Why B-scan sampling density matter?}
It is interesting to discuss the relationship between the channel numbers of input stacked BP data and the migration performance. In common sense, small spacing between consecutive measurements would lead to a high-performance migration result (i.e., a sharper, brighter, and more focused target point in the energy map). Still, it also brings a costly computation problem when processing a large amount of data. Therefore, how to balance the measurement sampling density and migration result is worth investigating.

Given raw B-scan data, we extract BP data with different number of channels, such as $64$, $128$, $256$, $128+64$, and $256+128$ as shown in Table~\ref{Table:channel_comparison}. In this way, we can distinguish the GPR imaging performances among all input types. Specifically, the metrics we used for performance evaluation are \textit{Mean Square Error (MSE)}, \textit{Signal-to-Noise-Ratio (SNR / dB)} and \textit{Structural Similarity Index (SSMI)}. Note that the lower \textit{MSE} value is, the better performance it present, while the higher \textit{SSMI} and \textit{SNR} value are, the better performance they present. 

The results indicate that our current input resolution: $256+128+64$, gains the best performance on \textit{SNR}, and second best performance on \textit{MSE} and \textit{SSMI} compared with other resolutions of input data. Notice when the input channel number decreases to $64$, it will go beyond the MigrationNet's ability to learn spatial features from such a sparse input. We also choose to reserve the raw input data without doing any sampling process, which leads to the best performance, but with more computation and longer processing time as expected.

\begin{table}[t]
\centering
\caption{Performance Comparison between the joint loss and Cross Entropy loss in MigrationNet}
\label{table:loss}
\setlength{\tabcolsep}{3mm}{
\begin{tabular}{|c|c|c|c|c|c|c|c|c|c|c|}
\hline
\hline
Metrics &\multicolumn{2}{|c|}{Joint Loss} &\multicolumn{2}{|c|}{Cross Entropy Loss}\\ 
\hline
\hline
Mean IoU & \multicolumn{2}{|c|}{\textbf{89.97}} &\multicolumn{2}{|c|}{{87.65}}\\
\hline
pixel Acc \%  & \multicolumn{2}{|c|}{\textbf{95.70}} &\multicolumn{2}{|c|}{{94.65}}\\
\hline
E\_distance & \multicolumn{2}{|c|}{\textbf{35.5809}} &\multicolumn{2}{|c|}{{38.9276}}\\
\hline
MSE  & \multicolumn{2}{|c|}{\textbf{661.1313}} &\multicolumn{2}{|c|}{{764.5629}}\\
\hline
SSMI  & \multicolumn{2}{|c|}{\textbf{0.9413}} &\multicolumn{2}{|c|}{{0.9378}}\\
\hline
SNR / dB  & \multicolumn{2}{|c|}{\textbf{3.3066}} &\multicolumn{2}{|c|}{{5.0584}}\\
\hline
\end{tabular}}
\end{table}

\textbf{Why the structure similarity loss matters?}
To verify the effectiveness of our joint loss, we further provide a comparison study with or without structural similarity loss as demonstrated in Table~\ref{table:loss}. As shown in Table~\ref{table:loss}, our joint loss has a better performance than the single Cross-Entropy loss, which reveals that this hybrid loss design can help capture structure information with a clearer boundary.

\begin{figure*}
    \centering
    \includegraphics[width=0.9\textwidth]{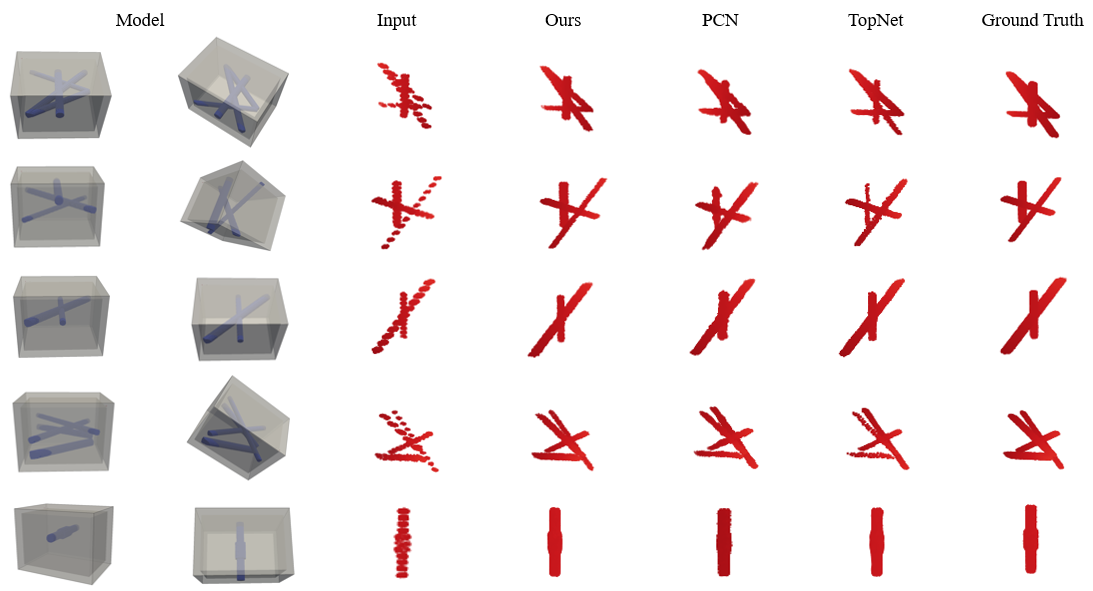}
    \caption{{Qualitative Comparison Results between GPRNet and baseline methods.} The comparison of completion results between other methods and our network. From left to right: the slab CAD model, input data, our method,  PCN\cite{yuan2018pcn}, TopNet\cite{tchapmi2019topnet}, and the ground truth. The results show our method could reconstruct a better 3D model for visualization.}
    \label{fig:pcn_result}
\end{figure*}

\subsection{Experimental Study of GPRNet}
\subsubsection{Effectiveness of GPRNet}
To evaluate the effectiveness of GPRNet, we compare with baseline methods such as PCN and TopNet\cite{tchapmi2019topnet} as shown in Table~\ref{Table:baseline} and Figure~\ref{fig:pcn_result}. In particular, we use three evaluation metrics for the quantitative effectiveness comparison in Table~\ref{Table:baseline}: Chamfer Distance (\textit{CD}), Earth Mover's Distance (\textit{EMD}) \cite{achlioptas2018learning} and $\mathcal{L}_{1}$ distance. Note that \textit{CD} indicates the average squared distance between two points; \textit{EMD} represents the average distance between corresponding points; $\mathcal{L}_{1}$ denotes the average distance from each point cloud to the centroid point in a point cloud set. Table~\ref{Table:baseline} indicates that our proposed method outperforms other methods in all the three evaluation metrics. Compared with the PCN and TopNet, GPRNet gains $5.9\%$ less and $7.2\%$ less chamfer distance respectively. Note that we up scale all the metrics by $10^3$ for better perception. 

In addition, the qualitative comparison result between GPRNet, PCN, and TopNet is depicted in Figure~\ref{fig:pcn_result}. Compared to PCN and TopNet, our proposed method obtains a better qualitative performance on different model structures where utility pipes have different radii embedded with different angles, depth and positions. Moreover, GPRNet can present the fine details of the object structure, such as a pipe with a joint at the center as shown at the bottom of Figure~\ref{fig:pcn_result}.
Based on the quantitative and qualitative results, we can conclude that our method outperforms the other methods in spatial continuity and shape accuracy level. 


\begin{table}[!th]
\caption{Quantitative Effectiveness Comparison results between GPRNet and other baselines.}
\label{table:weight}
\setlength{\tabcolsep}{4mm}{
\begin{center}
\begin{tabular}{|c|c|c|c|c|c|c|c|c|}
\hline
\hline
&\multicolumn{2}{|c|}{GPRNet} &\multicolumn{2}{|c|}{PCN} &\multicolumn{2}{|c|}{TopNet}\\ 
\hline
\hline
CD & \multicolumn{2}{|c|}{\textbf{6.328}}&\multicolumn{2}{|c|}{6.725} &  \multicolumn{2}{|c|}{{6.821}} \\
\hline
EMD & \multicolumn{2}{|c|}{\textbf{6.536}}&\multicolumn{2}{|c|}{6.827} &  \multicolumn{2}{|c|}{{7.173}} \\
\hline
$\mathcal{L}_{1}$ & \multicolumn{2}{|c|}{\textbf{2.016}} &\multicolumn{2}{|c|}{2.430} &  \multicolumn{2}{|c|}{{2.621}} \\
\hline
\end{tabular}
\end{center}}
\label{Table:baseline}
\end{table}

\begin{figure}
    \centering
    \includegraphics[width=0.45\textwidth]{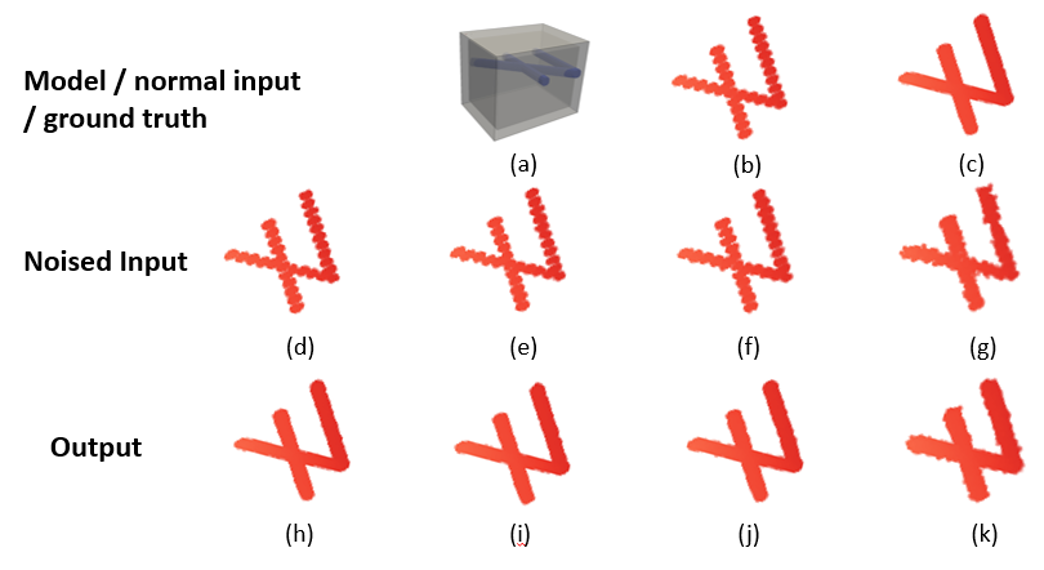}
    \caption{{Qualitative Noise Robustness results on GPRNet.} Pictures (a)-(c) indicate the model, raw input, and ground truth, respectively. The second line represents the Gaussian-white-noised input, where noise variance are $0.05$, $0.1$, $0.2$, $0.1$ respectively. Pictures (d)-(g) represent the noise input, while pictures (h)-(k) in the last line demonstrate the output of the models using GPRNet.}
    \label{fig:noise_comparision}
\end{figure}

\begin{table}[!th]
\caption{Evaluation Performance Comparison with various number of input point cloud.}
\label{table:incompletness}
\setlength{\tabcolsep}{1mm}{
\begin{center}
\begin{tabular}{|c|c|c|c|c|c|c|c|c|}
\hline
\hline
Sampling Number of Input Point Cloud &\multicolumn{2}{|c|}{CD} &\multicolumn{2}{|c|}{EMD} &\multicolumn{2}{|c|}{$\mathcal{L}_{1}$}\\ 
\hline
\hline
1000 & \multicolumn{2}{|c|}{7.264}&\multicolumn{2}{|c|}{7.498} &  \multicolumn{2}{|c|}{{2.629}} \\
\hline
\textbf{1500} & \multicolumn{2}{|c|}{6.328}&\multicolumn{2}{|c|}{6.536} &  \multicolumn{2}{|c|}{{2.016}} \\
\hline
2000 & \multicolumn{2}{|c|}{6.174} &\multicolumn{2}{|c|}{6.236} &  \multicolumn{2}{|c|}{{1.823}} \\
\hline
3000 & \multicolumn{2}{|c|}{5.524} &\multicolumn{2}{|c|}{5.563} &  \multicolumn{2}{|c|}{{1.585}} \\
\hline
4000 & \multicolumn{2}{|c|}{4.773} &\multicolumn{2}{|c|}{4.925} &  \multicolumn{2}{|c|}{{1.430}} \\
\hline
\end{tabular}
\end{center}}
\end{table}

\begin{table*}[htbp]
\caption{Noise Robustness Evaluation between GPRNet and baselines with three metrics.}
\label{table:NosieComparision}
\setlength{\tabcolsep}{1.5mm}{
\begin{center}
\begin{tabular}{|c|c|c|c|c|c|c|c|c|c|c|c|c|c|c|c|c|c|c|c|c|}
\hline
\hline
& \multicolumn{6}{|c|}{GPRNet} & \multicolumn{6}{|c|}{PCN} & \multicolumn{6}{|c|}{TopNet }\\ 
\hline
& \multicolumn{2}{|c|}{CD } & \multicolumn{2}{|c|}{EMD } & \multicolumn{2}{|c|}{$\mathcal{L}_{1}$ distance} & \multicolumn{2}{|c|}{CD } & \multicolumn{2}{|c|}{EMD } & \multicolumn{2}{|c|}{$\mathcal{L}_{1}$ distance}& \multicolumn{2}{|c|}{CD } & \multicolumn{2}{|c|}{EMD } & \multicolumn{2}{|c|}{$\mathcal{L}_{1}$ distance}\\ 
\hline
\hline
Variance \& Noise density = 0.01  &  \multicolumn{2}{|c|}{\textbf{6.419}}  &  \multicolumn{2}{|c|}{\textbf{6.628}}& \multicolumn{2}{|c|}{\textbf{2.287}} &  \multicolumn{2}{|c|}{6.901} &  \multicolumn{2}{|c|}{7.266}  &  \multicolumn{2}{|c|}{2.619} &  \multicolumn{2}{|c|}{6.894}&  \multicolumn{2}{|c|}{7.024}&  \multicolumn{2}{|c|}{2.5498}\\
Variance \& Noise density = 0.05 & \multicolumn{2}{|c|}{\textbf{7.722}} &  \multicolumn{2}{|c|}{\textbf{8.124}}& \multicolumn{2}{|c|}{\textbf{2.565}} &  \multicolumn{2}{|c|}{7.965} &  \multicolumn{2}{|c|}{8.313} &  \multicolumn{2}{|c|}{2.702} &  \multicolumn{2}{|c|}{8.248}&  \multicolumn{2}{|c|}{8.480}&  \multicolumn{2}{|c|}{2.7973}   \\
Variance \& Noise density = 0.1 &  \multicolumn{2}{|c|}{\textbf{7.774}} &  \multicolumn{2}{|c|}{\textbf{8.068}}&  \multicolumn{2}{|c|}{\textbf{2.601}} &  \multicolumn{2}{|c|}{8.141} &  \multicolumn{2}{|c|}{8.458} &  \multicolumn{2}{|c|}{2.783} &  \multicolumn{2}{|c|}{8.489}&  \multicolumn{2}{|c|}{8.517}&  \multicolumn{2}{|c|}{2.857}  \\
Variance \& Noise density = 0.2 & \multicolumn{2}{|c|}{\textbf{8.069}} &  \multicolumn{2}{|c|}{\textbf{8.369}} & \multicolumn{2}{|c|}{{3.046}} &  \multicolumn{2}{|c|}{8.495} &  \multicolumn{2}{|c|}{8.901} &  \multicolumn{2}{|c|}{\textbf{2.956}} &  \multicolumn{2}{|c|}{8.662}&  \multicolumn{2}{|c|}{8.956}&  \multicolumn{2}{|c|}{3.130}   \\
\hline
\end{tabular}
\end{center}
}
\end{table*}

\begin{figure*}
    \centering
    \subfigure[Conventional 2D migration result of CCNY test bed.]{
        \includegraphics[width=0.3\textwidth]{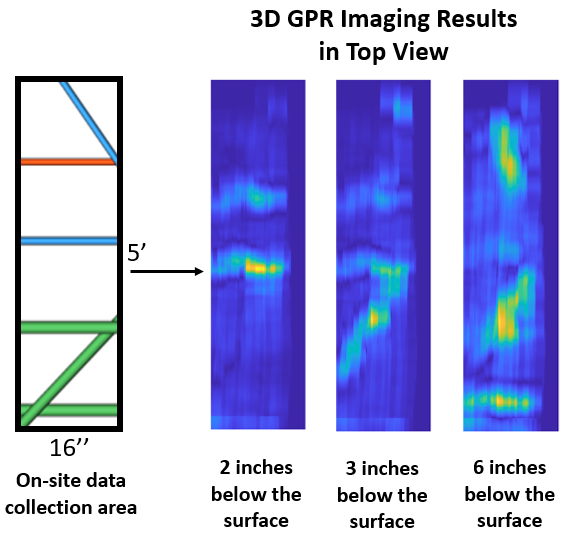}
    }
    \label{fig:2d}
    \subfigure[3D model of conventional migration method.]{
    	\includegraphics[width=0.3\textwidth]{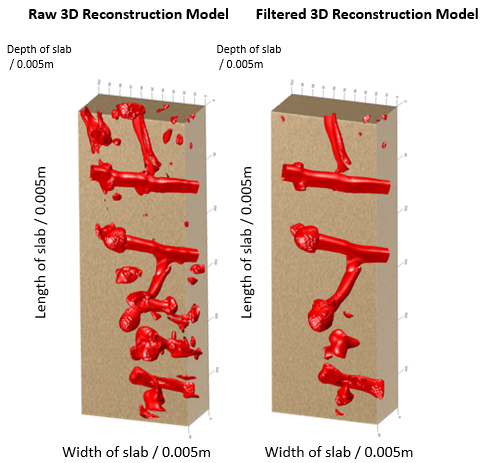}
    }
    \label{fig:3d}
    \subfigure[Point cloud based 3D model, where the color indicates the depth of the pipes.]{
        \includegraphics[width=0.3\textwidth]{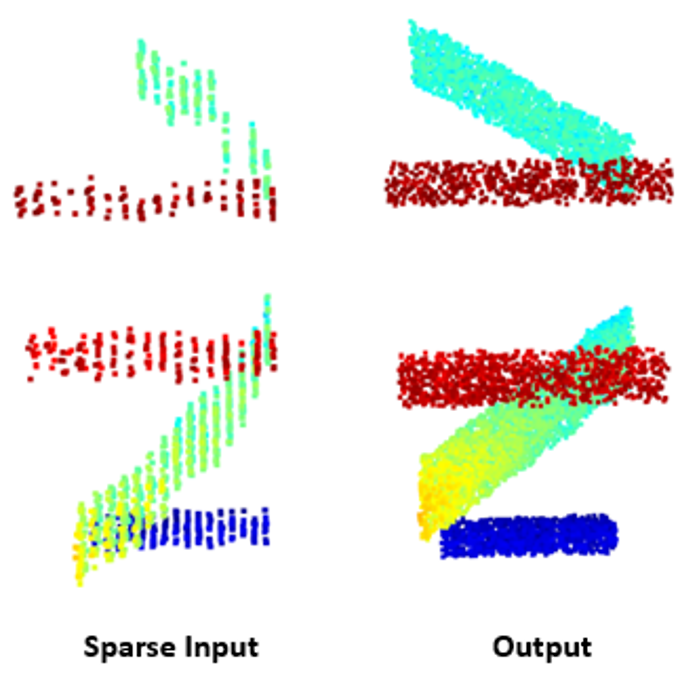}
    }
    \label{fig:3d_new}
    \caption{An illustration of GPR field 3D model results. (a) indicates the conventional migration 2D results for our collected field data at a concrete slab. (b) shows both the raw and filtered 3D model of the conventional migration result, while (c) shows the reconstruction point cloud model using GPRNet. }
    \label{fig:field_test}
\end{figure*}

Furthermore, we analyze the effectiveness of our network with different sampling numbers in input point clouds, where the number $N$ varies among $1000$, $1500$, $2000$, $3000$ and $4000$. In Table~\ref{table:incompletness}, we notice that the performance of GPRNet increases when the input number of point clouds increases. That is to say, the more point clouds sampled in the input, the less complicated the reconstruction task is. This experiment enlightens us to balance the sampling number of input point clouds and the performance of the reconstruction point cloud model. Thus, we decide to use $1500$ as the number of sparse input point cloud in all other studies, because it requires less collected B-scan data and computation time while it is still can achieve a relatively good reconstruction performance.

\subsubsection{Noise Robustness of GPRNet}
To evaluate the effectiveness of our method under different sensor noise levels, we perturbed the input sparse point cloud with multiple Gaussian white noise levels as shown in Figure~\ref{fig:noise_comparision}, where the standard deviations are 0.01, 0.05, 0.1, and 0.2, respectively. We further perform a quantitative study on noise comparisons among GPRNet, PCN, and TopNet with different metrics: \textit{CD}, \textit{EMD}, and $\mathcal{L}_{1}$ distance as illustrated in Table~\ref{table:NosieComparision}. We could conclude our proposed method gains higher robustness against the noise in comparison with PCN and TopNet.

\subsubsection{Field Test Model Reconstruction Comparison}
This section compares the effectiveness of the 3D reconstruction model between the proposed GPRNet and conventional migration method with the field data collected on the CCNY testbed. As illustrated in Figure~\ref{fig:field_test}, the black window region in Figure~\ref{fig:field_test}(a) indicates the data collection area while the three 2D images demonstrates migration results from the top view. Note that the red pipe in Figure~\ref{fig:field_test}(a) could not be recovered, the reason is that its depth is out of the GPR detection range (See Figure~\ref{fig:concrete_slab} for concrete slab details). Furthermore, we also illustrate raw and filtered 3D model generated by conventional migration methods in Figure~\ref{fig:field_test}(b), the deployed filter is Hessian filter \cite{pereira20203}. Due to the limitation of the conventional migration method, the noise data is hard to be cleaned out and differentiated from the raw GPR data, which causes the filtered 3D model is still hard to be recognized by normal GPR users. At last, Figure~\ref{fig:field_test}(c) illustrates the reconstructed point cloud model using GPRNet, where the color represents the depth of the pipe, the colder the color is, the deeper the reconstructed pipe buried. Note that in field test, the positioning accuracy would affect the distribution of the sparse input, which would introduce the noise in reconstituted point cloud model. However, under the supervision learning of the ground truth, the model could cover the uneven distributed area and fill up with point clouds, to reveal the real 3D model of the target. As we can see, compared with the traditional migration method, our method only requires a sparse input and further generate a fine and continuous output 3D model of underground pipes. It facilitates the GPR users to understand the complex raw GPR B-scan data. In addition to the better performance,  the data-collection time is also significantly reduced by using the proposed GPR-based robotic inspection platform. Without the robotic data collection platform, the inspector has to push the GPR device to follow exactly pre-marked grid lines, brings the GPR device back to the start points, marks the scanned points and takes the notes. This whole process would take around $30$ minutes to scan the black box area shown in Figure~\ref{fig:field_test}(a). However, our robotic-based data collection only needs less than three minutes to automatically scan the same area in free motion, which provides a more efficient way for GPR-based construction survey.

\section{Conclusion}
\label{section:conclusion}

This paper presents a robotic inspection system consisting of an Omni-directional robot and GPR post-processing software to automate the GPR data collection process and reconstruct 3D model of underground utilities for construction survey. Our omnei-directional robot allows the GPR device to move forward, backward, and sideways in a fast and swift manner. We propose a low-cost solution for vision-based accurate positioning, localization and mapping. By tagging the GPR position information in a synchronized way, it enables the robot to scan the surface in free-motion trajectory and facilitates high-resolution 3D GPR imaging. It eliminates the time, hassle, and cost to laying out grid lines on flat terrain and reducing the hassle to closely follow the grid lines and the note-taken time to record the linear motion trajectory in X-Y directions in current GPR data collection process. In addition, we propose a DNN-based method for 3D GPR imaging which contains two modules: MigrationNet and GPRNet. We evaluate the performance and validate the feasibility of our innovative method in the experimental studies. It demonstrates that our robotic inspection system makes the GPR data collection much easier by enabling the automatic scan of the flat surface in free-motion trajectories with minimal human intervention. By using synthetic data and real GPR data with ground truth value in our qualitative and quantitative experiments, it demonstrates that our 3D GPR imaging software can produce a fine 3D model of underground utilities. We are consulting with field engineers to collect large amount of field GPR data for training and testing to further increase the robustness of our method. The concrete slab GPR dataset is released publicly to research communities.

\section*{Acknowledgment}
Financial support for this study was provided by NSF grant IIP-1915721, and by the U.S. Department of Transportation, Office of the Assistant Secretary for Research and Technology (USDOT/OST-R) under Grant No. 69A3551747126 through INSPIRE University Transportation Center (http://inspire-utc.mst.edu) at Missouri University of Science and Technology. The views, opinions, findings and conclusions reflected in this publication are solely those of the authors and do not represent the official policy or position of the USDOT/OST-R, or any State or other entity. J Xiao has significant financial interest in InnovBot LLC, a company involved in R\&D and commercialization of the technology.

\ifCLASSOPTIONcaptionsoff
  \newpage
\fi

\bibliographystyle{IEEEtran}
\bibliography{IEEEabrv,Bibliography}

\begin{IEEEbiography}[{\includegraphics[width=1in,height=1.25in,clip,keepaspectratio]{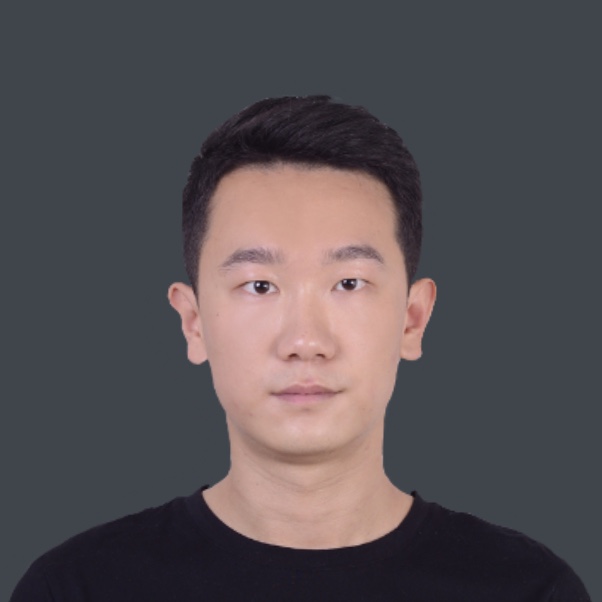}}]{Jinglun Feng} is currently pursuing the Ph.D. degree in electrical engineering with City College of New York, CUNY, New York, NY, USA. He received his B.S. degree in Electrical Engineering from Shandong Jianzhu University in 2015, and his M.S. degree in control engineering from the Shandong University in 2018. His research interests are in 3d vision, deep learning, object pose estimation, sensor fusion and intelligent inspection for robotics applications.
\end{IEEEbiography}

\begin{IEEEbiography}[{\includegraphics[width=1in,height=1.25in,clip,keepaspectratio]{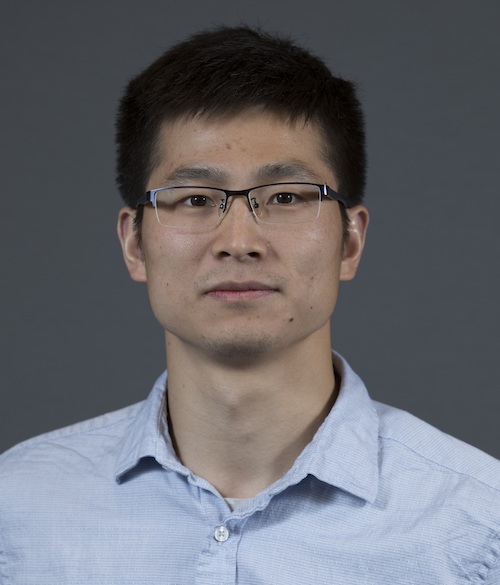}}]{Liang Yang} received his B.S. degree from Shenyang Aerospace University, Shenyang, China in 2012, and PhD degree in electronics engineering from the City College of New York (CUNY City College) in 2019, and PhD degree in pattern recognition and intelligent system from University of Chinese Academy of Sciences in 2019. He joined APPLE as a senior 3D computer vision researcher in 2019, and currently working on 3D visual perception and understanding. His research interests cover motion and path planning, 3D perception and understanding, visual SLAM and reconstruction, multi-sensor fusion, and robotic control.
\end{IEEEbiography}

\begin{IEEEbiography}[{\includegraphics[width=1in,height=1.25in,clip,keepaspectratio]{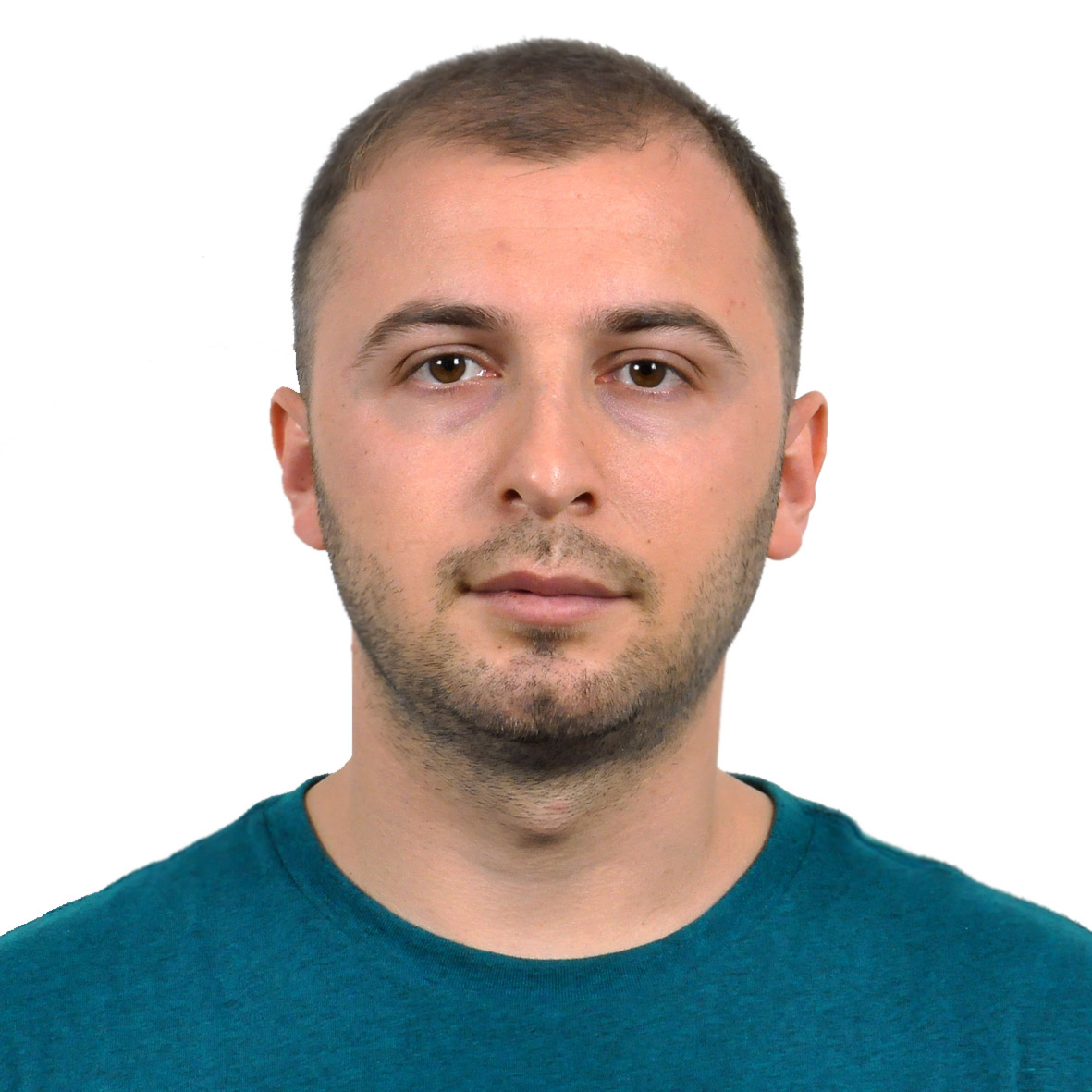}}]{Ejup Hoxha} is a Ph.D. student at The City College, City University of New York. He received his B.S. degree in Electrical and Computer Engineering with focus in control theory and robotics from University of Prishtina, and his M.S. degree in Computer Engineering from The City College, City University of New York. He worked several years in industry as a control engineer, SCADA developer and software developer. He is currently working at the CCNY Robotics Lab as researcher with a focus in robotics, control, visual SLAM, deep learning, sensor fusion, and acoustic NDT.
\end{IEEEbiography}

\begin{IEEEbiography}[{\includegraphics[width=1in,height=1.25in,clip,keepaspectratio]{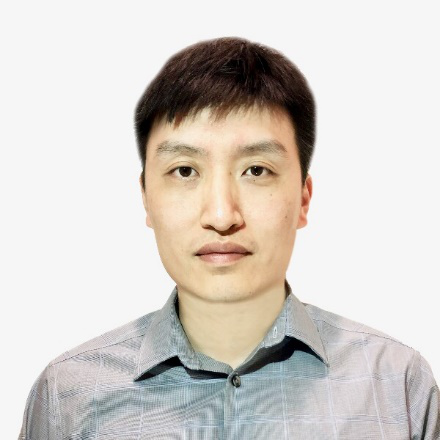}}]{Biao Jiang}
was born in Jiangsu, China. He received a Master in Interdisciplinary Studies in 2010, and a PhD in Electrical Engineering in 2013 from the City College of the City University of New York (CUNY), USA. He currently works at the Robotics Lab of City College of New York as a research scientist. His research interests include computer vision, SLAM and wireless communication.
\end{IEEEbiography}

\begin{IEEEbiography}[{\includegraphics[width=1in,height=1.25in,clip,keepaspectratio]{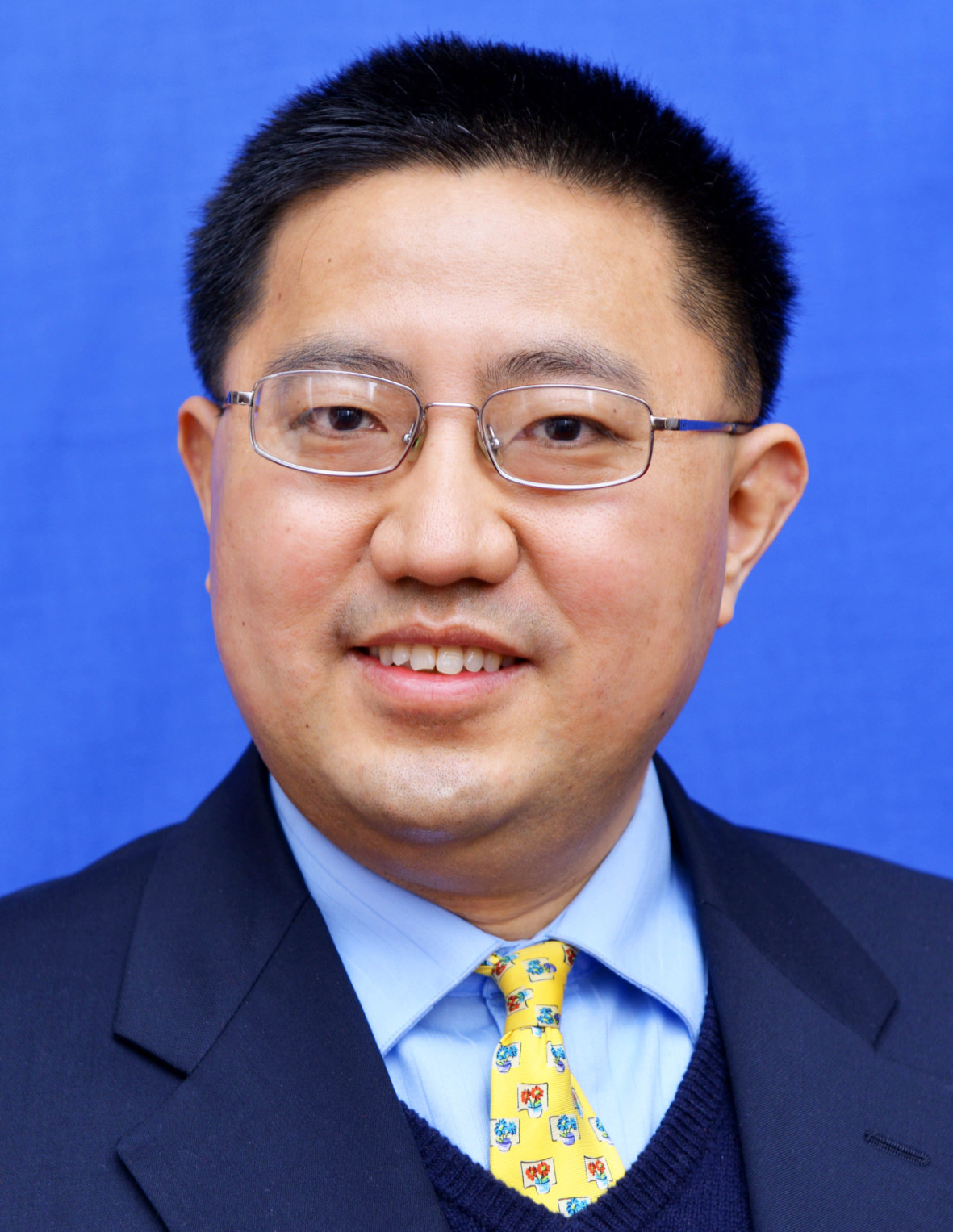}}]{Jizhong Xiao}
is a Professor of Electrical Engineering at the City College of New York (CCNY\/CUNY City College) and a doctoral faculty member of the Ph.D. program in Computer Science at CUNY Graduate Center. He received his Ph.D. degree from Michigan State University in 2002, M.E. degree from Nanyang Technological University, Singapore in 1999, M.S, and B.S. degrees from the East China Institute of Technology, Nanjing, China, in 1993 and 1990, respectively. He started the robotics research program at CCNY in 2002 as the founding director of CCNY Robotics Lab. His current research interests include robotics and control, cyber\-physical systems, autonomous navigation and 3D simultaneous localization and mapping (SLAM), real-time and embedded computing, assistive technology, multi\-agent systems and swarm robotics. He has published more than 160 research articles in peer reviewed journal and conferences. He received the U.S. National Science Foundation CAREER Award in 2007, the CCNY Outstanding Mentor Award in 2011, and the Humboldt Research Fellowship for Experienced Researchers from the Alexander von Humboldt Foundation, Germany, from 2013 to 2015. He is a senior member of IEEE.
\end{IEEEbiography}

\end{document}